%% file: paper.tex
\documentclass[twoside,11pt]{article}

\usepackage[preprint,abbrvbib]{jmlr2e}
\setcitestyle{numbers,square,comma}

\usepackage[USenglish]{babel}
\usepackage{doi}
\usepackage{microtype}
\usepackage[T1]{fontenc}
\usepackage{booktabs}
\usepackage{amsmath}
\usepackage[capitalize,noabbrev]{cleveref}

\usepackage[dvipsnames]{xcolor}
\hypersetup{colorlinks=true,linkcolor=MidnightBlue,citecolor=MidnightBlue,urlcolor=MidnightBlue}

\makeatletter
\long\def\@makecaption#1#2{%
  \vskip 10pt
  \noindent #1: #2\par
}
\makeatother

\title{Does Explanation Correctness Matter? Linking Computational XAI Evaluation to Human Understanding}
\ShortHeadings{Does Explanation Correctness Matter?}{Baer et al.}
\firstpageno{1}

\begin{document}

\author{
\name Gregor Baer$^{1}$,
\name Chao Zhang$^{2}$,
\name Isel Grau$^{1}$,
\name Pieter Van Gorp$^{1}$ \\
\addr $^{1}$ Information Systems Group, Eindhoven University of Technology \\
\addr $^{2}$ Human-Technology Interaction Group, Eindhoven University of Technology
}

\maketitle

\begin{abstract}
Explainable AI (XAI) methods are commonly evaluated using functional correctness metrics, also called faithfulness or fidelity, which estimate how closely an explanation reflects the model's reasoning. Higher correctness is assumed to produce better human understanding, but this link has not been tested with controlled levels. We conducted a user study ($N=200$) that manipulated explanation correctness at four levels (100\%, 85\%, 70\%, 55\%) in a synthetic time series classification task where participants could not rely on domain knowledge or visual intuition. Explanations were constructed to resemble feature attributions, with correctness defined against a known ground truth rather than estimated from a trained model. Participants predicted a simulated AI's decisions from these explanations (forward simulation), and we used their forward simulation accuracy as a proxy for understanding. Correctness affected forward simulation accuracy, but not at every level: accuracy dropped at 70\% and 55\% relative to 100\%, with no further reduction below 70\% and no conclusive difference between 85\% and 100\%. Lower correctness reduced how many participants predicted accurately rather than lowering accuracy uniformly, and even with fully correct explanations, only a subset of participants predicted well above chance. In exploratory analyses, some participants predicted decisions accurately but described the wrong pattern when asked what the AI used, and self-reported ratings correlated with forward simulation accuracy only when explanations were fully correct. These findings suggest that, for feature-attribution-style explanations, not all differences in functional correctness translate to differences in human understanding, highlighting the need to validate functional metrics against human outcomes.
\end{abstract}


\section{Introduction}\label{h:intro}

Explainable AI (XAI) methods generate explanations for AI model decisions to support human understanding and oversight. The field has grown rapidly~\citep{nauta.etal_2023_anecdotal,piculin.etal_2025_position}, but evaluating these explanations remains an open challenge~\citep{longo.etal_2024_explainable}. Two broad approaches exist~\citep{doshi-velez.kim_2018_considerations,speith_2026_review}: functional (human-independent) evaluation scores explanations against computational criteria such as robustness or correctness, while human-based evaluation measures human outcomes such as understanding or human-AI team performance~\citep{rong.etal_2024_humancentered}. Most work in the machine learning and computer science communities relies on functional evaluation, with user studies being comparatively rare~\citep{nauta.etal_2023_anecdotal,piculin.etal_2025_position,rong.etal_2024_humancentered,suh.etal_2025_fewer}, partly because they are costly and time-consuming.

The prevalence of functional evaluation implies that the field optimizes XAI methods for computational metrics that may or may not relate to human outcomes, implicitly assuming that functional metrics serve as meaningful proxies for human understanding. One widely used functional property is correctness (also called faithfulness or fidelity), which estimates how accurately a post-hoc explanation reflects the model's reasoning. 
It is routinely used both as a design criterion for developing new explanation methods and as a yardstick for evaluating existing ones~\citep[e.g.,][]{cheng.etal_2026_uncertainty,grau.napoles_2024_sparsenessoptimized,gruensteidl.kirrane_2026_evaluation,khan.etal_2024_faithful,turbe.etal_2023_evaluation}.
The assumption is that explanations scoring poorly on correctness will mislead users or fail to support understanding. 

In practice, users receive explanations whose correctness they cannot verify. Because post-hoc XAI methods approximate complex black-box models with simpler representations, they rarely capture the model's reasoning in full. They can also disagree with one another~\citep{krishna.etal_2024_disagreement} and do not always reflect the underlying model~\citep{adebayo.etal_2018_sanity,kindermans.etal_2019_unreliability}. Correctness can vary between methods and from one explanation to the next. It is also not one metric but a family of them, operationalized in many ways and organized under competing evaluation taxonomies~\citep{nauta.etal_2023_anecdotal,speith_2026_review}, with no standard that applies across settings~\citep{longo.etal_2024_explainable}. Users are therefore left interpreting explanations of unknown and varying correctness.

The question this raises is whether those variations impact the user at all.
Functional correctness may differ in ways that leave human understanding unaffected. Differences may be hard to notice, or they may occur in parts of the explanation that users do not attend to or do not treat as relevant for their decisions. People may also learn useful mental models from imperfect explanations as long as the explanations preserve enough useful information, with effects emerging only once inaccuracies exceed a tolerance threshold. If either mechanism holds, optimizing for small differences in functional correctness may yield limited gains in human outcomes.

A small number of studies have investigated how explanation properties relate to human outcomes, but none provide a controlled experimental test of how correctness at varying levels affects understanding. Existing work has treated correctness as binary~\citep{cabitza.etal_2024_explanations,morrison.etal_2024_impact,spitzer.etal_2025_dont}, relied on correlational rather than experimental designs~\citep{colin.etal_2022_what,kares.etal_2026_are,nguyen.etal_2021_effectiveness}, and all used image tasks where human intuition can compensate for poor explanations~\citep{cabitza.etal_2024_explanations,colin.etal_2022_what,kares.etal_2026_are,morrison.etal_2024_impact,nguyen.etal_2021_effectiveness,spitzer.etal_2025_dont}. Therefore, while linking functional metrics to human outcomes has been identified as a priority for advancing XAI~\citep{piculin.etal_2025_position}, the question remains open.

To address this gap, we conducted a between-participants user study ($N=200$) that tested whether explanation correctness, treated as a continuum rather than as correct or incorrect, affected human understanding of AI decision-making.
We operationalized understanding as forward simulation accuracy: how well participants, after a training phase in which they saw the AI's classifications, predict the AI's decisions from the explanations alone.
In a task with no domain knowledge or visual intuition to fall back on, predicting decisions above chance should require some grasp of the decision rule. This follows an abilities-based account of understanding, in which demonstrating an ability such as prediction counts as evidence of understanding without any single ability capturing it in full~\citep{speith.etal_2024_conceptualizing}.

Because we simulated both the data and the classifier, the ground truth explanation was known, so we could set explanation correctness at exact levels rather than estimating it from a trained model. We used time series rather than images, so participants could not fall back on visual intuition or domain knowledge that might otherwise confound results~\citep{chen.etal_2023_machine}. 
We presented explanations as highlighted regions of the time series, following the format of post-hoc feature attribution methods.
This allowed us to test how understanding, as measured by forward simulation accuracy, was affected by varying levels of explanation correctness.
Together, these design choices provide a first experimental test of whether, and to what degree, functional correctness affects human understanding of AI explanations.

We found that explanation correctness affected participants' ability to predict the AI's decisions, but not at every level. Forward simulation accuracy was lower at 70\% and 55\% correctness than with fully correct explanations, but reducing correctness from 70\% to 55\% did not reduce it further, consistent with a threshold effect. Our data do not locate that threshold precisely, and the comparison between the two highest correctness levels was inconclusive. Fully correct explanations helped only a subset of participants predict the AI's decisions accurately. Some participants who predicted the AI's decisions accurately still described the wrong pattern when asked what the AI used, so two abilities associated with understanding, predicting decisions and explaining the rule behind them, did not always agree. Self-reported ratings also correlated with forward simulation accuracy only when explanations were fully correct, raising questions about when subjective measures are informative.

The remainder of this paper is organized as follows. We first review related work on feature attribution evaluation, correctness metrics, and existing evidence linking correctness to human outcomes. We then describe the study design, including the synthetic time series task, the correctness manipulation, and the forward simulation measure. Next, we present results from the planned hypothesis tests alongside exploratory analyses of the progression of accuracy throughout the experiment, self-reported ratings, and the relationship between subjective and objective measures, followed by a general discussion of the findings. We close with limitations, conclusions, and directions for future work.

\section{Related Work}\label{h:related-work}

\subsection{Feature attributions and evaluation}\label{h:related-work-xai-evaluation}

Post-hoc XAI methods explain AI models after training. Among the various types of post-hoc explanations, including rules, concept-based explanations, and counterfactuals~\citep{dwivedi.etal_2023_explainable,guidotti.etal_2018_survey}, feature attributions are among the most common. They assign an importance score to each input feature, such as a pixel, time point, or word, indicating how much it contributed to the model's prediction. 
Well-known examples include LIME~\citep{ribeiro.etal_2016_why} and SHAP~\citep{lundberg.lee_2017_unified}, alongside other gradient-based~\citep{bach.etal_2015_pixelwise,shrikumar.etal_2017_learning,simonyan.etal_2014_deep,smilkov.etal_2017_smoothgrad,sundararajan.etal_2017_axiomatic} and perturbation-based methods~\citep{fong.vedaldi_2017_interpretable,zeiler.fergus_2014_visualizing}. These attributions are typically visualized as heatmaps, often called saliency maps in the image domain, that highlight which parts of the input drive the prediction. The implicit promise is that showing users \emph{what} the model attends to will help them understand \emph{how} it decides. 

Evaluating explanations is difficult because, unlike prediction tasks, there is no ground truth for what the correct explanation should be. \citet{doshi-velez.kim_2018_considerations} distinguish three evaluation approaches: application-grounded evaluation tests explanations with domain experts on real tasks~\citep[e.g.,][]{chanda.etal_2025_dermatologistlike,chanda.etal_2024_dermatologistlike}; human-grounded evaluation uses simplified proxy tasks to measure properties like understanding, trust, or human-AI collaboration performance~\citep[for an overview, see][]{rong.etal_2024_humancentered}; and functionally-grounded evaluation bypasses humans entirely and scores explanations with computational metrics. In practice, the XAI community relies heavily on functional evaluation, while user studies remain less common~\citep{nauta.etal_2023_anecdotal,piculin.etal_2025_position}.

Even when user studies are conducted, the evidence on whether explanations help users is mixed. Several studies show that explanations can increase reliance on AI decisions regardless of whether the AI is correct~\citep{bansal.etal_2021_does,kaur.etal_2020_interpreting} and that users' perceived understanding often exceeds what they can demonstrate on objective tasks~\citep{chromik.etal_2021_think}. Effects on objective understanding are similarly uncertain: they vary across explanation types and datasets, with some explanation formats improving users' prediction accuracy and others showing no benefit or even negative associations~\citep{colin.etal_2022_what,hase.bansal_2020_evaluating}. At the same time, explanations can sometimes reduce overreliance and improve decisions~\citep{bucinca.etal_2025_contrastive,dejong.etal_2025_cognitive,vasconcelos.etal_2023_explanations}. 

This mixed picture raises a question that sits between the functional and human-based evaluation paradigms: if we improve the functional correctness of an explanation, does human understanding actually improve? Whether this promise holds depends on two links: the explanation must accurately reflect the model's reasoning (correctness), and that accuracy must translate to human understanding. 
The first link is the subject of a large functional evaluation literature, the second is largely untested.
Terminology across this literature varies considerably, with many competing taxonomies~\citep{speith_2026_review}. \cref{fig:concept-relations} therefore maps the two evaluation paradigms, the constructs they target, and how we operationalize each in this study, and serves as a reference for the terminology used throughout this paper. We review related work from both evaluation paradigms in turn.

\input{figt_concept_relations}

\subsection{Measuring correctness}\label{h:related-work-measuring-correctness}

Among the functional properties that researchers evaluate, such as compactness, completeness, and others~\citep[for a taxonomy, see][]{nauta.etal_2023_anecdotal}, correctness is one of the most widely evaluated. Correctness, also called faithfulness and sometimes fidelity, captures how accurately an explanation reflects the model's reasoning. The rationale for prioritizing it is that an explanation that misrepresents the model may mislead users regardless of how sparse or stable it is. 
In practice, the boundaries between these constructs are not always sharp when operationalized: widely used perturbation metrics measure a blend of correctness, compactness, and completeness rather than correctness alone~\citep[see Figure~7 in][]{nauta.etal_2023_anecdotal}.
Taxonomies also vary, using similar terms with slightly different meanings and grouping the same metrics under different properties~\citep{speith_2026_review}.
We use the term correctness broadly, to refer to whether an explanation faithfully reflects what drives the model, encompassing both families of metrics described below.

Perturbation analysis~\citep{samek.etal_2017_evaluating} measures correctness by sequentially removing or altering features in order of their attributed importance. The assumption is that if an attribution correctly identifies what matters, removing those features should degrade model predictions. Variants include deletion and insertion metrics~\citep{petsiuk.etal_2018_rise}, remove-and-retrain approaches~\citep{hooker.etal_2019_benchmark}, and methods that combine different orders of perturbation~\citep{bluecher.etal_2024_decoupling,schulz.etal_2020_restricting,turbe.etal_2023_evaluation}. 
Sensitivity- and perturbation-based analysis has a long history in machine learning that predates XAI. In the context of XAI evaluation, it was first applied to assess explanations in the image domain and was later extended to other data domains including time series~\citep[e.g.,][]{schlegel.etal_2019_rigorous,simic.etal_2025_comprehensive}.
However, these metrics have known limitations: they can be sensitive to how features are perturbed~\citep{bluecher.etal_2024_decoupling,simic.etal_2025_comprehensive}, yield different results across predicted classes~\citep{baer.etal_2025_classdependent,baer.etal_2026_why}, and different operationalizations can disagree with each other~\citep{hedstrom.etal_2023_metaevaluation}.

Localization-based metrics offer an alternative when ground truth feature locations are known. Instead of perturbing inputs, they measure the overlap between the attribution and the known discriminative region, for instance, through precision, recall, or intersection over union. Metrics tailored to attributions include the pointing game~\citep{zhang.etal_2018_topdown}, AUC-based measures~\citep{arias-duart.etal_2022_focus}, and relevance mass and rank accuracy~\citep{arras.etal_2022_clevrxai}. Since real-world datasets rarely provide ground truth, this approach typically requires synthetic data where discriminative features are controlled by design~\citep[e.g.,][]{baer.etal_2026_why,ismail.etal_2020_benchmarking,nguyen.ifrim_2025_tshap,turbe.etal_2023_evaluation}. The common assumption is that a model achieving high accuracy on such data must rely on the intended features. However, high accuracy alone does not guarantee this, and evaluating against a fixed ground truth has been criticized for being insensitive to the actual model being explained~\citep{rawal.etal_2025_evaluating}.

Despite their different mechanics, both approaches aim to quantify the same underlying construct: how well an explanation captures what drives the model's predictions. Neither measures it directly. Each is an operationalization with its own assumptions and possible failures, and the two families can disagree about the same explanation ~\citep[e.g.,][]{baer.etal_2026_why}. Both are nonetheless widely used, and researchers routinely report higher scores as evidence of better explanation quality, implicitly assuming that these improvements benefit users. Our study tests this assumption directly. Our correctness manipulation aligns with the localization logic, as we control the ground truth and measure temporal overlap with a known feature location, but the question we ask, i.e., whether correctness affects human understanding, applies regardless of how correctness is operationalized.

\subsection{Correctness and human understanding}\label{h:related-work-correctness-understanding}

Several studies have examined whether explanation correctness affects users, though most have focused on trust rather than understanding. Some studies found that explanations increased trust even when incorrect~\citep{kim.etal_2022_hive,lakkaraju.bastani_2020_how,sadeghi.etal_2024_explaining}, while others reported that nonsensical or low-correctness explanations could harm trust~\citep{papenmeier.etal_2019_how,wang.etal_2024_impact}. 
However, trust and understanding are distinct constructs: a user may trust AI more after seeing an explanation, even if they do not understand how it works. 
Conceptual models of the XAI explanation process place understanding upstream of downstream outcomes such as satisfaction and appropriate reliance~\citep{hoffman.etal_2023_measures,langer.etal_2021_what,speith_2026_review}, which is why we study it directly rather than through trust or team performance.
We focus on studies that examined how correctness relates to human understanding, grouping them into experimental and correlational studies.

Three studies have directly manipulated explanation correctness. \citet{morrison.etal_2024_impact} asked participants to classify birds with AI support, providing explanations that were either correct or deliberately incorrect or misleading, meaning logically inconsistent with the presented stimuli. Incorrect explanations decreased classification accuracy, particularly for non-experts.
~\citet{spitzer.etal_2025_dont} extended this to an architectural style classification task with text-based explanations, finding that incorrect explanations impaired both human-AI team performance and participants' ability to perform the task independently after collaboration.
~\citet{cabitza.etal_2024_explanations} reported similar patterns using logic puzzle tasks with inconsistent text explanations. 
While these findings show that clearly wrong explanations can harm performance, all studies measured human-AI collaboration accuracy rather than model understanding, so their effects could stem from changes in trust or reliance rather than comprehension. They also used tasks where humans have strong visual or logical intuitions and can independently detect errors, and treated correctness as binary, which does not reflect the continuous scale on which functional metrics are measured. In many practical settings, users work with less interpretable data such as time series or tabular inputs, where they cannot fall back on intuition. In these settings, the primary question shifts from whether incorrect explanations erode trust to whether explanations can create understanding in the first place.

A second line of work has taken a correlational approach, exposing participants to explanations from different XAI methods and checking whether functional metric scores correspond to human performance. 
~\citet{nguyen.etal_2021_effectiveness} found no relationship between functional metrics and human-AI team performance for image classification, though participants frequently succeeded despite poor heatmap localization, suggesting that the tasks were solvable through visual recognition alone.
~\citet{colin.etal_2022_what} used a forward simulation design closer to ours: participants learned from saliency map explanations with the AI's output shown, and then predicted the AI's output for new inputs from the explanations alone.
Across multiple image tasks, they found no clear link between correctness metrics and prediction accuracy. For some datasets, more correct explanations were negatively associated with understanding, possibly reflecting confounding by explanation complexity. 
~\citet{kares.etal_2026_are} compared three saliency map methods on ImageNet across self-reported measures, objective proxy tasks for human understanding, and functional metrics, finding that the three evaluation families did not agree on which method performed the best. Since these studies compared different XAI methods that also varied in visual appearance, sparsity, and complexity, observed differences in human outcomes cannot be attributed to correctness alone. Moreover, the differences in functional metric scores between the compared methods were often small and not specifically targeted, which may be insufficient to produce detectable effects in human performance.

Neither line of work establishes whether functional correctness, treated as a continuous property, affects human understanding of AI decision-making. The experimental studies tested binary contrasts between correct and deliberately misleading explanations rather than varying degrees of correctness; the correlational studies lacked experimental control over correctness levels; and all relied on image tasks where human intuition can compensate for poor explanations. ~\citet{chen.etal_2023_machine} showed formally that human intuitions can influence how people derive understanding from explanations, making image-based tasks less suitable for isolating the effect of explanation correctness. These gaps call for an experimental approach that manipulates correctness at controlled levels in a domain where participants must genuinely rely on explanations to understand the AI.

\section{Method}

\subsection{Study Design Overview}

\input{figt_study_design}

We conducted a user study in which participants learned to predict how an AI classifies instances based on explanations, a task known as \emph{forward simulation}~\citep{doshi-velez.kim_2018_considerations,lipton_2018_mythos}.
Participants viewed time-series plots with an overlaid heatmap explanation; the heatmap resembled a feature-attribution visualization. 
Each instance's classification by the simulated AI was fixed when the stimulus was generated (\cref{h:methods-stimuli}), and the heatmap indicated which time points this classification was based on.
Participants predicted whether the AI would label each instance as Class~A or Class~B.
We manipulated the correctness of these explanations as a single between-participants factor at four levels: 100\%, 85\%, 70\%, and 55\% temporal overlap between the displayed heatmap and the ground truth discriminative feature region.

\cref{fig:study-design} illustrates the study flow. After providing informed consent, participants received instructions and completed a comprehension check. They were then randomly assigned to one of four conditions.
They completed 20 training trials, each followed by feedback showing the AI's actual classification, and then 20 test trials in which this classification was withheld.
The primary outcome was forward simulation accuracy on the test trials: the proportion of predictions matching the AI's classification. The study concluded with a survey collecting subjective ratings and open-ended responses. The study was implemented with lab.js~\citep{henninger.etal_2022_labjs} and hosted at our university premises.
We preregistered hypotheses, sample size, and an analysis plan prior to data collection. The anonymized dataset, analysis scripts, and stimuli generation code are deposited on Zenodo.\footnote{Preregistration: \url{https://osf.io/y49ne/overview?view_only=aa561514157b48ad858f6ae18b22e6f0}. Data and code: \url{https://doi.org/10.5281/zenodo.19330434}.}
The ethical review board of our university approved this study (ERB-527).

\subsection{Participants}

\input{tab_demographics}

We recruited 200 participants through Prolific who had at least a bachelor's degree, an approval rate of 98\% or higher, and current residence in the UK or US. Participants received \pounds3.00 for an estimated 20 minutes of work. 
We targeted 50 participants per condition. We estimated the population standard deviation conservatively at $SD = 0.173$, derived from a uniform distribution over plausible accuracy bounds $[0.35, 0.95]$. With this estimate, $n = 50$ provides 95\% confidence intervals of $\pm 0.05$ around condition means and 80\% power to detect accuracy differences of 8.6 percentage points in one-tailed Welch's $t$-tests ($\alpha = 0.05$). For non-inferiority tests with margin $\Delta = 0.10$ and true effects near zero, the same sample provides 89\% power.

We randomly assigned participants to conditions, balancing across the four groups. We preregistered exclusion criteria prior to data collection; no participants met any. 
Two participants in the 55\% condition voluntarily withdrew after assignment and were replaced, creating a slight imbalance.
Final sample sizes were $n=51$ (100\% and 85\% correct), $n=50$ (70\%), and $n=48$ (55\%), totaling 200 completed submissions. We analyzed all 200 completed submissions. Median completion time was 12 minutes and 18 seconds. 
\cref{tab:demographics} reports participant demographics by condition.

\subsection{Experimental Procedure}

\input{figt_trial_screen}

\cref{fig:trial-screen} shows a trial as seen by participants. Axes were unlabeled and the heatmap legend used qualitative labels (low to high importance) rather than numeric values, since the time series were normalized and specific values carried no meaningful interpretation for the task. Participants completed a comprehension check before entering the experiment; those who failed twice were screened out before condition assignment. During training, participants received feedback after each prediction showing the AI's actual classification, allowing them to learn the relationship between the explanation and the AI's decisions. The test phase used the same task structure but without feedback. Presentation order was randomized independently for each participant in both phases.

The post-task survey collected two open-ended responses: participants described what pattern they thought the AI uses to distinguish classes, and how their approach changed during training. Participants also rated five statements about explanation quality and reported demographic information.

\subsection{Stimuli and Correctness Manipulation}\label{h:methods-stimuli}

We operationalized correctness as the temporal overlap between the displayed heatmap and the true discriminative feature, following localization-based evaluation in the XAI literature~\citep{nauta.etal_2023_anecdotal}. Manipulating correctness at controlled levels requires knowing the ground truth explanation, which real-world AI models and XAI methods do not provide. This also rules out the common perturbation-based correctness metrics, which do not have a ground truth and are additionally sensitive to parameter choices~\citep{baer.etal_2025_classdependent,bluecher.etal_2024_decoupling,schlegel.keim_2023_deep,simic.etal_2025_comprehensive}, making them difficult to control experimentally. We instead constructed synthetic time series with a known decision boundary and presented them as outputs of an AI classifier. No actual AI model was used; the task decision boundary and the supposed model decision boundary were identical by design. This eliminates confounds that would arise with a real pipeline, where the observed relationship between correctness and understanding would be entangled with model accuracy, explanation correctness, and metric reliability.

Univariate time series of length 200 were generated with additive composition by using the \texttt{xaitimesynth} package~\citep{baer_2026_xaitimesynth}, following previous work~\citep{baer.etal_2026_why}. Each time series $\mathbf{x} = \mathbf{n} + \mathbf{f}$ combined a random walk base signal $\mathbf{n}$ (step size 0.15) with a feature component $\mathbf{f}$ containing the class-discriminating feature within a designated window and zeros elsewhere. The feature was either a peak with amplitude $1$ (Class~A) or a valley with amplitude $-1$ (Class~B), spanned 15\% of the series length (30 time points), and was placed at a random temporal location. We calibrated these parameters so that the discriminative feature region is not visually obvious without the heatmap: the random walk produces peaks and valleys of similar magnitude, so participants must rely on the explanation to identify which region drives the classification. \cref{fig:data-generation} shows the data generation approach with these parameters for two instances from each class.

\input{figt_data_generation}

The ground truth explanation is a binary array marking where the feature component was present. 
Because the feature component determines the class label, and the AI's classification was identical to that label by design, the ground truth explanation indicates why each instance received its classification. 
All heatmaps were generated from this ground truth explanation, so each heatmap explains its instance's classification, with the correctness level determining how accurately it localizes the relevant region.
Even with fully correct explanations, the heatmap indicated where the class-discriminating feature was located, but not whether it was a peak or a valley (Class~A or Class~B). Because the feature location was randomized independently of the class, participants could not infer the AI's classification from the heatmap alone and had to examine the time series in the highlighted region.

\cref{fig:stimuli-creation} illustrates the explanation generation process. To create realistic heatmaps, we corrupted 30\% of randomly selected timesteps with uniform noise from $[0, 0.8]$, decreasing values in the feature region and increasing values outside it. The result was then smoothed with a Gaussian filter ($\sigma=1$). To create less correct explanations, we spatially displaced the ground truth region before generating the heatmap, with direction randomized with equal probability. In the 85\% condition, for example, 85\% of the highlighted area covered the discriminative feature while 15\% covered an adjacent non-informative region. Visual properties of the heatmaps, including color intensity, smoothness, and overall appearance, remained constant across conditions. The only difference between conditions was the spatial position of the ground truth region from which the heatmap was generated, which was our correctness manipulation.

\input{figt_stimuli_creation}

\subsection{Measurements}\label{h:methods-measurements}

\textbf{Primary measure: forward simulation accuracy.}
The primary outcome measure, specified in our preregistration, was forward simulation accuracy: the proportion of the 20 test trials on which a participant correctly predicted the AI's decision from the explanation.
As in other forward simulation studies, the AI's classification was withheld during the test phase, since displaying it would allow participants to read off the answer rather than infer it from the explanation.
We additionally recorded trial-level accuracy during the 20 training trials, which we use in an exploratory analysis of the learning process.
Both experiment phases contained equal numbers of Class~A and Class~B instances, so chance performance was 0.50.
Because the random walk of the time series can partly mask the feature, some instances may be ambiguous even when the rule is applied perfectly, so the attainable maximum may lie below 1.

Forward simulation is commonly used as a behavioral proxy for measuring participants' understanding of AI decision-making~\citep[e.g.,][]{alqaraawi.etal_2020_evaluating,chromik.etal_2021_think,colin.etal_2022_what,hase.bansal_2020_evaluating}. 
It rests on the assumption that to predict and \emph{forward simulate} the AI's decisions above chance in a task where no domain knowledge or visual intuition is available, participants must have acquired some understanding of the underlying decision rule.
Such proxy tasks do not necessarily translate to human-AI team performance~\citep{bucinca.etal_2020_proxy}, but the ability to predict the AI's decisions is a minimal requirement: if explanations cannot support it in a controlled setting where no domain knowledge or visual intuition is available, they are unlikely to improve decision-making in practice.
Forward simulation thus differs from human-AI decision-making paradigms, in which participants see the AI's output alongside the explanation and the outcome of interest is reliance or joint human-AI performance~\citep{rong.etal_2024_humancentered}.
In the conceptualization of \citet{speith.etal_2024_conceptualizing}, forward simulation targets the \emph{predict} ability, one of several abilities through which understanding can be operationalized and measured.

\textbf{Secondary measure: open-ended responses.}
Participants answered two open-ended questions after the test phase. The first asked them to describe the pattern or feature they believed the AI used to distinguish Class~A from Class~B time series, which asks them to verbalize the decision rule rather than predict individual decisions. We treat the resulting verbalized decision rules as an exploratory analysis, letting us analyse whether participants who predict the AI's decisions can also verbalize the rule behind them.

Because the data were simulated, the underlying ground-truth rule was known, so we coded each response deductively into three categories: \emph{correct}, when the response named the localized peaks or troughs that define the classes; \emph{incorrect}, when it named another characteristic such as trend, variance, or the location of the highlighted regions; and \emph{none}, when the participant reported no pattern or described only the heatmap rather than the underlying data (e.g., stating that the AI ``focused on the hot areas'').
The coding reflects whether participants named the class-discriminative feature, not whether they accurately described what the heatmap highlighted. In the lower correctness conditions, the heatmap partly highlighted non-discriminative features such as trends, so a participant could correctly describe what they saw in the explanation and still be coded as \emph{incorrect} because those features did not distinguish the classes. This mirrors applied XAI settings, where the goal is for users to identify the feature driving the model's decisions, but an incorrect explanation can lead them to form a coherent understanding of the wrong feature.

Two coders independently coded all 200 responses using the same guidelines, reaching initial agreement on 173 responses (86.5\%; Cohen's $\kappa = .745$). They then discussed and resolved the 27 disagreements to reach consensus. \cref{h:verbalized-rules} reports results based on these consensus codings.

During coding, we noticed that some participants described only the heatmap rather than any feature of the underlying time series. We therefore also coded this as a binary variable (heatmap-only response) using the same procedure, reaching initial agreement on 194 of 200 responses (97\%; Cohen's $\kappa = .841$).
In the second question, participants described how their approach to predicting the AI's decisions changed over the training trials. We summarize these responses briefly in the results and release them with the dataset.

\textbf{Secondary measure: self-reported ratings.}
Participants rated five statements on Likert scales, adapted from the Explanation Satisfaction Scale~\citep{hoffman.etal_2023_measures}, which we analyzed individually. 
The Explanation Satisfaction Scale is widely used in XAI but has not been psychometrically validated, so we treat it as an exploratory secondary measure. We dropped two of the seven items that assume the participant knows how to use the model or how accurate it is, neither of which applies to our forward simulation task, and report the remaining items individually so readers can interpret them directly.
These items capture \emph{perceived} explanation quality, including perceived understanding: they record participants' subjective judgments rather than a demonstrated ability. Whether perceived and demonstrated understanding correspond is an empirical question, which we address in our exploratory analyses.

\input{tab_test_results}

\subsection{Hypotheses and Statistical Tests}

We predicted that human understanding of AI decision-making, as measured by forward simulation accuracy, decreases as explanation correctness decreases. To characterize this relationship, we tested five planned comparisons. Baseline comparisons quantified the total cost of degradation relative to perfect explanations: 85\% correct vs.\ 100\% (H1), 70\% vs.\ 100\% (H2), and 55\% vs.\ 100\% (H3). Adjacent comparisons tested whether each step of degradation produces additional reductions in accuracy: 70\% vs.\ 85\% (H4) and 55\% vs.\ 70\% (H5). 
Together, these allowed us to examine whether understanding degraded gradually with correctness or remained stable until a certain level.

For each comparison, we conducted two complementary one-tailed \textit{t}-tests. The first tested whether the lower-correctness condition reduced accuracy relative to the higher-correctness condition, that is, whether the mean difference was significantly below zero. However, failing to detect a reduction does not establish that the effect is absent or negligibly small. We therefore paired it with a second test (known as a non-inferiority test) that asked whether we could rule out reductions larger than $\Delta = 0.10$, corresponding to 2 additional errors per 20 test trials. If the non-inferiority null is rejected, any reduction is estimated to be smaller than $\Delta$, even if the first test finds a statistically significant difference.

Combining both tests produces four interpretive outcomes: (1)~\emph{meaningful reduction}: a significant reduction is detected that cannot be ruled out to exceed $\Delta$; (2)~\emph{small but detectable}: a significant reduction is detected but confirmed to fall below $\Delta$; (3)~\emph{no meaningful difference}: no significant reduction is detected and effects as large as $\Delta$ can be ruled out; or (4)~\emph{inconclusive}: a meaningful reduction can neither be confirmed nor ruled out. We set $\alpha = 0.05$ for all tests without correction for multiple comparisons, as each comparison addresses a distinct question about whether improvements at a specific correctness level benefit human understanding.

\section{Results and Discussion}\label{h:results-discussion}

\subsection{Forward simulation accuracy and correctness}\label{h:fsim}

\cref{tab:hypothesis-tests} summarizes the results of all five planned comparisons.
Mean forward simulation accuracy was .690 (SD = .237) in the 100\% correct condition, .625 (SD = .179) in 85\%, .544 (SD = .131) in 70\%, and .525 (SD = .125) in 55\%.

Correctness affected forward simulation accuracy, but not at every level. We call a reduction meaningful when we cannot rule out that it exceeds $\Delta = .10$, which corresponds to 2 additional errors per 20 test trials.
The comparison between 85\% and 100\% correctness (H1) was inconclusive. Accuracy was .066 lower at 85\%, 95\% CI $[-.148, .017]$. Neither test rejected ($p_0 = .059$, $p_\Delta = .206$), so the interval covers reductions larger than $\Delta$ as well as no reduction at all.
Accuracy was .146 lower at 70\% and .165 lower at 55\% than at 100\% (H2, H3). Both reductions were significant, both estimates exceed $\Delta$, and neither test against $\Delta$ rejected, so both count as meaningful reductions. The adjacent step from 85\% to 70\% (H4) was also a meaningful reduction, with accuracy .081 lower, 95\% CI $[-.142, -.019]$. Here the estimate falls below $\Delta$, but the interval does not rule out larger reductions.
The step from 70\% to 55\% (H5) showed no meaningful difference. Accuracy was .019 lower, no reduction was detected ($p_0 = .232$), and the test against $\Delta$ rejected ($p_\Delta = .001$), ruling out reductions of .10 or larger.

\input{figt_forward_sim_accuracy}

\cref{fig:forward-sim-accuracy} shows the forward simulation accuracy distributions by condition.
In the 100\% condition, accuracy was bimodal: some participants reached high accuracy while others remained near chance.
As correctness decreased, high accuracy became rare: the 85\% condition still had participants with high forward simulation accuracy, whereas in the 70\% and 55\% conditions no participant exceeded $.75$.
Degraded correctness therefore did not lower accuracy uniformly. Instead, it reduced how many participants reached high accuracy while the rest stayed near chance. 
Why some participants failed to predict the AI's decisions even with fully correct explanations remains unclear. 
The high variance in the 100\% and 85\% conditions likely contributed to H1 being inconclusive despite a mean difference of $.066$.
Variance also decreased as correctness dropped (SD = .237, .179, .131, .125), consistent with this picture: with fewer high scores, accuracy clustered near chance.

Overall, correctness affected forward simulation accuracy, but not at every level. 
Forward simulation accuracy dropped significantly at 70\% and 55\%, plateaued from 70\% to 55\%, and the drop between 85\% and 70\% was confirmed. Together, these results point to a threshold pattern rather than a linear decline. The distributions suggest that below the threshold, high accuracy largely disappears, rather than accuracy across the condition simply being lower. Whether mild degradation (85\%) already decreased demonstrated understanding remains unresolved, though four discrete levels cannot fully characterize the functional form.

\subsection{Exploratory results}

\subsubsection{Learning process}

\input{figt_learning_curves}

The forward simulation accuracy reported so far comes from the 20 test trials, which participants completed without feedback. To see how that accuracy developed, we turn to the 20 training trials that came before it.
During training, feedback after each prediction let participants test and refine how the explanations related to the AI's decisions, so learning that relationship should show up as forward simulation accuracy rising across trials.

\cref{fig:learning-curves} plots cumulative forward simulation accuracy across the training trials, averaged across participants within each condition. 
The curves therefore show group trends, not individual trajectories.
In the 100\% correct condition, the lower bound of the 95\% confidence interval cleared chance by trial 7 and stayed above it. In the other three conditions, the confidence interval included chance throughout training. The 85\% condition edged above the 70\% and 55\% conditions late in training but not far enough to separate from chance, and the 70\% and 55\% conditions stayed near chance across all trials, mirroring their near-chance test-phase accuracy. Only fully correct explanations moved the group mean above chance during training, and this separation emerged within the first several trials, showing that above-chance prediction of the AI's decisions appeared early where it appeared at all.

\subsubsection{Verbalized decision rule}\label{h:verbalized-rules}

Alongside forward simulation accuracy, we collected open-ended responses. Forward simulation accuracy measures how well participants predicted the AI's decisions. The open-ended question measures whether participants verbalized the feature they think drives those decisions. If a participant predicted the AI's decisions well because they picked up on the real pattern, they should be able to name that pattern, assuming they can put it into words. Comparing the two shows how far predicting and verbalizing agree.

\input{tab_verbalized_rule.tex}

After the test phase, we asked participants to describe the pattern or feature they thought the AI used to distinguish Class~A from Class~B. We coded each response against the known ground-truth rule into three categories (correct, incorrect, none). The coding procedure is described in \cref{h:methods-measurements}. \cref{tab:verbalized_rule} reports the coded counts by correctness condition.

The counts follow the same direction as forward simulation accuracy in \cref{h:fsim}. Correct verbalizations concentrated in the higher correctness conditions: 14 and 17 participants identified the decision rule at 100\% and 85\%, against 4 and 2 at 70\% and 55\%. Incorrect verbalizations remained common across all conditions, and \emph{none} responses rose slightly as correctness dropped, from 6 to 9 and 9 to 13.

\input{figt_fsim_rule_relationship}

At the individual level, predicting accurately and verbalizing correctly did not always correspond to each other. \cref{fig:fsim-rule-relationship} shows the relationship between forward simulation accuracy and the verbalized decision rule for the 100\% and 85\% conditions, the only two with enough correct verbalizations to examine.
At 100\% correct, the link between verbalizing and predicting is relatively clear. Of the 14 participants who verbalized the rule correctly, 12 reached forward simulation accuracy $\geq .8$. Those who could verbalize the pattern thus also seem to be able to predict from it.
Correct verbalization was, however, not necessary for accurate prediction. Of the 23 participants with forward simulation accuracy $\geq .8$, only 12 verbalized the rule correctly, 9 verbalized it incorrectly, and 2 reported no pattern. Roughly half the participants who predicted the AI's decisions at $\geq .8$ named the rule, and the other half reached the same accuracy while describing something else or nothing at all.

At 85\% correct, the relationship between verbalization and forward simulation accuracy is less clear. Although 17 participants verbalized the rule correctly (more than at 100\%), only 8 of them reached accuracy $\geq .8$, while 9 scored below. Correct verbalization at this level corresponded less consistently to accurate prediction, and incorrect verbalizations spanned the full range of forward simulation accuracy in both conditions. The predict-verbalize link thus held at the group level but broke down for individual participants, and seemingly more so as correctness decreased.

This divergence could partly be the result of the difficulty of accurately verbalizing a decision rule in a few sentences, especially for an abstract task with unfamiliar data such as time series.
A participant who learned to predict from the pattern may still not be able to articulate what they learned. 
In the other direction, the few cases where participants verbalized the rule correctly but scored below $.8$ on forward simulation accuracy, may partly reflect noise in coding short free-text responses. 
For instance, a brief mention of ``peaks'' can be coded as correct without indicating full understanding of the rule. Both measures are imperfect, and we treat the verbalized rule as a secondary, exploratory measure that complements forward simulation accuracy rather than replacing it.

We also counted 23 participants in total (6, 1, 7, and 9 across the four conditions from highest to lowest correctness) whose open-ended responses referred exclusively to the heatmap or highlighted regions without describing any feature of the time series. Although the counts are small, they suggest that some participants engaged only with the visual overlay rather than the underlying data, which may partly explain the wide range of forward simulation accuracy observed even with fully correct explanations. For these participants, the task difficulty may have been less about explanation correctness and more about making sense of the explanation format itself.

A second open-ended question asked how participants' approach to predicting the AI's decisions changed over the training phase. Most reported trial and error, no change in strategy, or not finding a pattern, which largely reflects the training design rather than distinguishing between participants.

\subsubsection{Self-reported ratings}

\input{figt_explanation_ratings}

\cref{fig:explanation-ratings} shows self-reported explanation ratings across conditions, measured with five items adapted from the Explanation Satisfaction Scale~\citep{hoffman.etal_2023_measures}. Higher correctness conditions tended toward more positive ratings, but differences between conditions were modest. The items on detail and completeness received the least favorable ratings: even in the 100\% condition, fewer than 35\% of participants agreed that explanations seemed complete. Notably, self-reported understanding remained relatively high even in conditions where forward simulation accuracy was substantially lower: in the 70\% and 55\% conditions, 40\% and 37\% of participants agreed or strongly agreed that they understood how the AI works, compared to 59\% in the 100\% condition. Self-reported ratings thus declined more gradually across conditions than forward simulation accuracy (\cref{fig:forward-sim-accuracy}).

\subsubsection{Self-reports vs. forward simulation accuracy}

\input{figt_corr_ess_avg}

\cref{fig:correlation-ess-avg} shows that, averaged across conditions, the five self-reported items correlated strongly with each other ($r = .65$ to $.83$) but weakly with forward simulation accuracy ($r\leq.24$). The subjective ratings thus formed a consistent cluster that was largely separate from demonstrated understanding.

\input{figt_corr_ess_split}

However, splitting correlations by condition reveals that this weak average is partly misleading. \cref{fig:correlation-ess-split}A shows that in the 100\% condition, correlations between self-reports and forward simulation accuracy were weak to moderate ($r = .28$ to $.43$). In the remaining conditions, correlations mostly dropped toward zero, with two exceptions that likely reflect sampling variability: helpfulness reached $r = .25$ at 85\% correctness and $r = .22$ at 55\%. This drop was not proportional to the decline in forward simulation accuracy. The 85\% condition still included participants with high forward simulation accuracy (mean accuracy $.625$, $SD = .179$), yet correlations with self-reports were already near zero for most items. The link between self-reports and forward simulation accuracy thus seemed to break down earlier than the decline in forward simulation accuracy alone would suggest.

The 100\% condition was bimodal, with participants clustering at either high or near-chance forward simulation accuracy. We therefore asked whether the correlations reflected the condition as a whole or only the participants who predicted the AI's decisions accurately. We split the condition at a forward simulation accuracy of $.80$, which corresponds to 16 of the 20 test trials correct ($n = 23$ and $28$), and recomputed the self-report correlations within each half (\cref{fig:correlation-ess-split}B). At or above $.80$, self-reports correlated weakly to moderately with forward simulation accuracy ($r = .21$ to $.33$). Below $.80$, correlations were near zero ($r = -.15$ to $.10$). A split at the condition median ($.70$) gives a similar pattern.
This exploratory split should be interpreted cautiously: it conditions on the primary outcome itself, so neither half's correlation reflects the full relationship between self-reports and forward simulation accuracy, and both halves are small ($n = 23$ and $28$), making the estimates imprecise.
Even so, the within-condition and condition-level results agree: self-reports correlated with forward simulation accuracy only when explanations were fully correct and participants predicted the AI's decisions accurately. When either was missing, they did not.

\subsection{General discussion}

Three main findings emerge for the feature-attribution-style explanations we tested.
First, not all differences in functional correctness translated to differences in human understanding as measured by forward simulation accuracy. Degrading correctness from 100\% to 70\% or 55\% significantly reduced forward simulation accuracy. Accuracy at 55\% was no lower than at 70\%, and we could rule out reductions larger than $\Delta = .10$, which corresponds to 2 additional errors per 20 test trials. Further reductions in correctness therefore did not produce further reductions in accuracy, which suggests a threshold somewhere between 100\% and 70\% correctness.

We cannot say where in that range the threshold sits. The comparison between 100\% and 85\% was inconclusive in both directions: we detected no reduction in forward simulation accuracy, and the non-inferiority test did not rule out a reduction larger than $\Delta$ ($p_\Delta = .206$). The estimated reductions for the steps from 100\% to 85\% and from 85\% to 70\% were also close, .066 and .081, with overlapping confidence intervals. What separates them is whether the reduction reached significance. Differences of this size are plausible between XAI methods, and our data do not resolve whether they matter for users. Where current methods fall on this scale is itself unclear, since correctness values depend on which metric is used.

The distributional evidence supports the threshold interpretation. In the 70\% and 55\% conditions, forward simulation accuracy was unimodal and clustered around chance. In the 100\% condition it was bimodal (\cref{fig:forward-sim-accuracy}), with one group near chance and one well above it. Lower correctness appeared to reduce how many participants could accurately predict the AI's decisions, rather than uniformly shifting forward simulation accuracy downward.
The verbalized rules point the same way. About a quarter of participants named the ground-truth feature, localized peaks and troughs, at 100\% and 85\% correctness (14 of 51 and 17 of 51). Almost none did at 70\% and 55\% (4 of 50 and 2 of 48).
All conditions provided an explanation, so these comparisons are relative to fully correct explanations rather than to no explanation at all.

Second, fully correct explanations alone did not guarantee that participants could predict the AI's decision. Only a subset of participants reached high forward simulation accuracy (\cref{fig:forward-sim-accuracy}), and several of those who predicted the AI's decisions accurately named the wrong feature when asked to verbalize the decision rule. Our data cannot explain why the others stayed near chance. 
A correct explanation shows which part of the time series the model used, but participants still had to infer the rule themselves, and we designed the task so that domain knowledge and visual intuition could not help with that step.
\citet{kenny.etal_2021_explaining} distinguish users' mental models of the domain, the AI system, and the explanation strategy, and note that performance measures can reflect any of them. Our design removed the first, so forward simulation accuracy in our task likely reflects a mix of participants' mental models of the explanation and of the AI. Participants who predicted accurately but named the wrong feature may have learned how to use the explanation without forming an accurate model of the AI's decision rule, or may have been unable to put that rule into words. For the 23 participants across conditions who described only the heatmap, making sense of the explanation format itself may have been the problem.
Factors beyond correctness, such as how explanations are presented and how users are supported in interpreting them, may therefore be necessary for explanations to support understanding. 
For XAI evaluation, this implies that high correctness scores alone are not sufficient evidence that explanations will help users understand model behavior well enough to predict it.

Third, self-reported ratings correlated with forward simulation accuracy only when explanations were fully correct, and even then mainly among participants who predicted well. Averaged across conditions, the correlation between subjective ratings and forward simulation accuracy was weak, consistent with prior work that showed similar disconnects between subjective self-reports and demonstrated behavior~\citep{abbaspouronari.etal_2025_dynamics,bucinca.etal_2020_proxy,chromik.etal_2021_think,hase.bansal_2020_evaluating,papenmeier.etal_2019_how,wang.etal_2024_impact}. Exploratory analyses suggest that self-reported ratings may be least informative where they would matter most: even at 100\% correctness, they correlated with forward simulation accuracy among participants who already predicted well, but said little about accuracy among those who did not.

Taken together, these results suggest that for feature-attribution-style explanations, differences in functional correctness do not necessarily translate to differences in understanding, in the sense of being able to predict model behavior. In our task, a moderate loss of correctness reduced forward simulation accuracy, further losses changed little, and full correctness did not ensure high accuracy. Whether this pattern holds for other explanation types, data modalities, and correctness metrics remains to be tested, which is why we see value in validating functional metrics against human outcomes rather than assuming the link.

\section{Limitations}\label{h:limitations}

We used forward simulation accuracy as our measure of understanding. This assumes that, in a task with no domain knowledge or visual intuition to fall back on, predicting the AI's decisions above chance requires some understanding of the decision rule. That assumption has limits. Prediction can be seen as one component of understanding~\citep{speith.etal_2024_conceptualizing}, not a full account of a user's mental model. A participant might thus predict well without fully grasping the underlying decision ruled for instance by learning to associate the shape under the highlighted region with a class.
Our open-ended responses also point in this direction. More participants stated the correct rule under fully correct explanations than under degraded ones, yet, some who predicted decisions accurately still described the rule incorrectly. We interpret this carefully, because an abstract time-series pattern is hard to put into words, which can make coding the open-ended responses noisy. Forward simulation accuracy therefore captures demonstrated understanding, but only what a participant can predict, not all possible components of understanding.

We ran the study online and cannot fully verify participant effort. A comprehension check confirmed task understanding before the main task, but we cannot rule out careless responding by some participants.

Finally, our design favors experimental control over ecological validity, a typical tradeoff for controlled proxy tasks~\citep{doshi-velez.kim_2018_considerations}. 
Controlling correctness at precise levels requires a known ground truth, so we used synthetic time series with a known decision boundary rather than a real model and XAI pipeline. This control comes at a cost along several dimensions.
Our correctness manipulation displaced a single contiguous feature region in time, so correctness is operationalized as temporal overlap with the ground-truth region. Perturbation-based metrics are more common in evaluation practice, but they vary with class and parameter choices~\citep{baer.etal_2025_classdependent,baer.etal_2026_why} and cannot be set to specific levels. Whether the two notions capture the same property is an open question we do not test here.

Our simulated explanations were built to resemble real feature attributions, but real XAI output differs in complexity and noise, and real methods may produce error patterns we did not model, such as noisy or systematically biased attributions that could affect understanding differently than the displacement we studied. 
We also kept the decision rule simple: the simulated AI classified each series from the peaks and troughs in a single localized region, a pattern participants could plausibly learn to predict within a short session. It was not trivially easy, since only a subset reached high forward simulation accuracy even with fully correct explanations, but real AI systems are usually more complex. 
Every condition included an explanation, which lets us isolate the effect of correctness. We therefore cannot separate whether the near-chance performance at 70\% and 55\% reflects explanations that carry no usable information or explanations that draw attention away from the ground-truth feature.
These choices leave open how far the results extend. We tested only heatmap-style feature attributions for univariate time series classification, and other explanation formats, data modalities, or tasks may behave differently.

\section{Conclusion and Future Work}\label{h:conclusion}

We conducted a controlled experiment that manipulated the correctness of feature-attribution-style explanations at four levels and measured human understanding through forward simulation. Forward simulation accuracy was lower at 70\% and 55\% correctness than with fully correct explanations, reducing correctness below 70\% produced no further reduction, and the comparison between 85\% and 100\% was inconclusive. 
Fully correct explanations did not guarantee high forward simulation accuracy either, and self-reported ratings correlated with forward simulation accuracy only when explanations were fully correct. In exploratory analyses, some participants who predicted the AI's decisions accurately still described the wrong pattern when asked what the AI used, so predicting decisions and describing the rule behind them did not always agree. For feature-attribution-style explanations, these findings suggest that correctness scores alone are not sufficient proxies for how well users understand model behavior. This does not mean functional metrics lack value, but higher correctness scores should not be assumed to produce better understanding without empirical validation.

This matters because of how XAI methods are currently evaluated. Functional metrics are the most common way to assess them, while studies with users remain rare~\citep{nauta.etal_2023_anecdotal,piculin.etal_2025_position,rong.etal_2024_humancentered,suh.etal_2025_fewer}. Higher average functional scores are treated as better without much evidence about what they imply for the people the explanation methods are developed for in the first place. We tested one such metric against one human outcome for one explanation type and found that the two do not line up as cleanly as assumed. Whether other functional metrics fare better is an open question, and one that current evaluation practice rarely asks.

These results open several directions for future work. 
Testing more correctness levels between 100\% and 70\% would show where the threshold lies and resolve the comparison between 85\% and 100\%. Differences of this size are plausible between XAI methods, though where current methods fall is unclear. A second extension concerns what correctness is measured against. In our study, the simulated AI always decides correctly and follows the ground truth rule, so a correct explanation is right about the AI and right about the task at the same time. In practice, models are not always right. A model can learn a spurious pattern, and an explanation can describe that pattern accurately, which makes the explanation right about the model and wrong about the task. A future study could vary AI correctness and explanation correctness separately. This would show how the two interact: whether users notice that the AI follows a flawed reasoning when the explanation shows correctly shows that reasoning, and whether an explanation that matches the task rather than the AI helps or hurts human outcomes.

Correctness alone did not ensure that participants could predict the AI's decisions, so identifying other factors that support this kind of understanding is equally important. Domain expertise is one candidate. We deliberately chose an abstract task to prevent participants from relying on prior intuitions, but in practice, expertise could compensate for or amplify the effects of explanation correctness. Another candidate is how explanations are presented and how users are supported in interpreting them. A related open question is what participants who predicted accurately had actually learned. Some named the wrong feature, which could mean they learned to use the explanation without forming an accurate model of the AI, or that they could not put what they learned into words. A single open-ended question cannot separate these possibilities. Qualitative methods such as think-aloud protocols or follow-up interviews could show more directly how participants used the explanations to reach their predictions.

Explanation correctness may also moderate the link between self-reported ratings and forward simulation accuracy, since the two correlated when explanations were fully correct but not at lower correctness levels. If this pattern replicates, the disconnect between subjective and objective measures reported in prior work may depend on the explanations participants are given. More broadly, the experimental approach we used here, manipulating a functional property at controlled levels and measuring human outcomes, could be applied to other widely used metrics such as robustness or sparsity, and to other explanation types such as counterfactual or concept-based explanations, where correctness takes a different form and would need its own manipulation.

\section*{Declaration of Competing Interest}
The authors declare that they have no known competing financial interests or personal relationships that could have appeared to influence the work reported in this paper.

\section*{Data Availability}
The anonymized dataset, analysis scripts, and stimuli generation code are deposited on Zenodo at \url{https://doi.org/10.5281/zenodo.19330434}.

\section*{Acknowledgements} 
Funding: This work was supported by the European Union’s HORIZON Research and Innovation Program under grant agreement No. 101120657, project ENFIELD (European Lighthouse to Manifest Trustworthy and Green AI). 

\bibliography{references.bib}

\end{document}

%% file: figt_concept_relations.tex
\begin{figure*}[!b]
\centering
\includegraphics[width=0.8\textwidth]{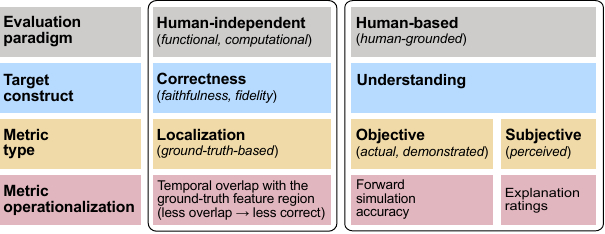}
\caption{From evaluation paradigms to metric operationalizations for correctness and understanding. The two XAI evaluation paradigms target different constructs: correctness describes the explanation, understanding describes the state of the user. Italics denote alternative names used in the literature; we use \emph{functional} evaluation throughout the paper. We manipulate correctness by generating heatmaps with controlled temporal overlap with a known ground-truth feature region (100/85/70/55\%), rather than estimating correctness from a trained model. We measure understanding objectively through forward simulation accuracy and subjectively through self-reported ratings.}
\label{fig:concept-relations}
\end{figure*}

%% file: figt_study_design.tex
\begin{figure*}[!b]
\includegraphics[width=\textwidth]{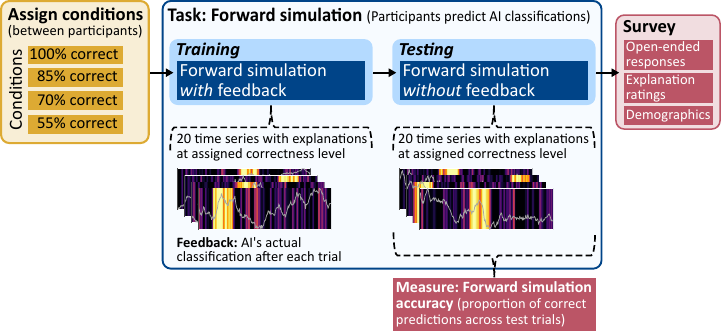}
\caption{Overview of the experimental design, showing condition assignment, training and test phases, and the post-task survey. During training, participants saw the AI's actual classification after each prediction. During testing, it was withheld to measure forward simulation accuracy.}
\label{fig:study-design}
\end{figure*}

%% file: tab_demographics.tex
\begin{table}[!htbp]
\caption{Participant demographics by correctness condition.}
\label{tab:demographics}
\centering
\small
\setlength{\tabcolsep}{4pt}
\begin{tabular*}{\columnwidth}{@{\extracolsep{\fill}} l r r r r r r r @{}}
\toprule
 & & \multicolumn{2}{c}{Age} & \multicolumn{2}{c}{Gender} & \multicolumn{2}{c}{Technical} \\
\cmidrule(lr){3-4} \cmidrule(lr){5-6} \cmidrule(lr){7-8}
Correctness & $n$ & Mean & SD & Man & Woman & Yes & No \\
\midrule
100\% & 51 & 41.7 & 12.5 & 25 & 26 & 16 & 35 \\
85\% & 51 & 42.3 & 10.6 & 29 & 22 & 19 & 32 \\
70\% & 50 & 44.7 & 11.4 & 29 & 20 & 18 & 32 \\
55\% & 48 & 41.8 & 11.5 & 28 & 19 & 13 & 35  \\
\midrule
Total & 200 & 42.6 & 11.5 & 111 & 87 & 134 & 66 \\
\bottomrule
\end{tabular*}

\raggedright
\smallskip
\footnotesize\textit{Note.} Gender excludes one non-binary and one undisclosed participant. Technical = self-reported technical background.
\end{table}

%% file: figt_trial_screen.tex
\begin{figure*}[!tb]
\centering
\includegraphics[width=0.8\textwidth]{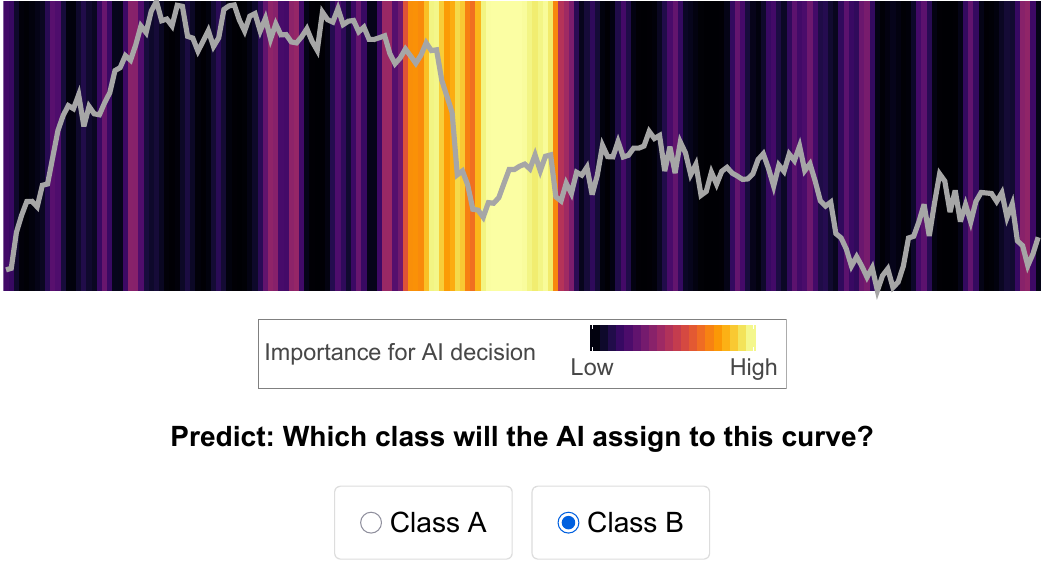}
\caption{Trial screen as shown to participants during both the training and test phases. Participants see a univariate time series curve overlaid with a heatmap indicating feature importance for the AI's decision. In the training phase, the next screen shows feedback with the AI's actual classification; in the test phase, no feedback is provided. The example shown here is from the 70\% correctness condition, where the heatmap partially overlaps with the true discriminative feature and partially covers an adjacent non-informative region.}
\label{fig:trial-screen}
\end{figure*}

%% file: figt_data_generation.tex
\begin{figure*}[!tb]
\includegraphics[width=\textwidth]{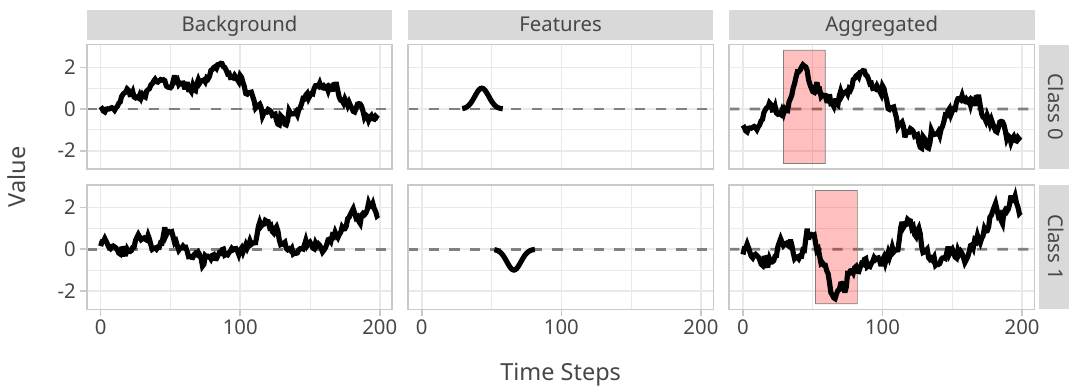}
\caption{Synthetic data generation. Each time series combined a random walk (left) with a class-discriminative feature (center): a peak (Class~A) or valley (Class~B) spanning 15\% of the series length. The aggregated signal (right) is the sum of both components, with the red region marking the ground truth feature location. The random walk produced peaks and valleys of similar magnitude, making the feature difficult to identify without an explanation.}
\label{fig:data-generation}
\end{figure*}

%% file: figt_stimuli_creation.tex
\begin{figure*}[!tb]
\centering
\includegraphics[width=\textwidth]{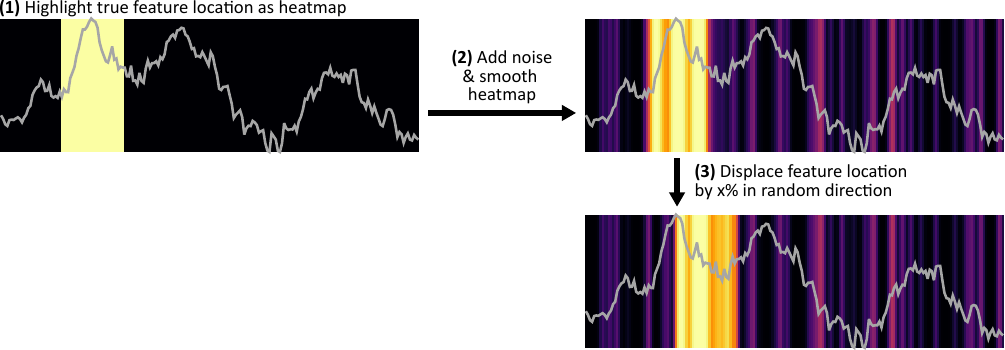}
\caption{Explanation generation and correctness manipulation. The true feature location was rendered as a heatmap, corrupted with noise, and smoothed to resemble a feature attribution explanation. Less correct explanations were created by spatially displacing the highlighted region. The example shown corresponds to a displacement of 45\%, leaving 55\% overlap with the true feature location (i.e., the 55\% correctness condition).}
\label{fig:stimuli-creation}
\end{figure*}

%% file: tab_test_results.tex
\begin{table*}[!htbp]
\caption{Effect of explanation correctness on forward simulation accuracy.}
\label{tab:hypothesis-tests}
\centering
\small
\setlength{\tabcolsep}{3pt}
\begin{tabular*}{\textwidth}{@{\extracolsep{\fill}} l l r l r r r r r r l @{}}
\toprule
 & Comparison & Diff. & 95\% CI & $d$ & \textit{df} & $t_0$ & $p_0$ & $t_\Delta$ & $p_\Delta$ & Interpretation \\
\midrule
\multicolumn{11}{@{}l}{\textit{Baseline comparisons (vs.\ 100\% correctness)}} \\[2pt]
H1 & 85\% vs 100\% & $-.066$ & [-.148, .017] & $-.313$ & 93.02 & $-1.579$ & .059 & $0.825$ & .206 & Inconclusive \\
H2 & 70\% vs 100\% & $-.146$ & [-.222, -.070] & $-.761$ & 78.31 & $-3.844$ & 1.2e-4 & $-1.215$ & .886 & Meaningful reduction \\
H3 & 55\% vs 100\% & $-.165$ & [-.240, -.090] & $-.865$ & 76.64 & $-4.375$ & 1.9e-5 & $-1.727$ & .956 & Meaningful reduction \\
\midrule
\multicolumn{11}{@{}l}{\textit{Adjacent comparisons}} \\[2pt]
H4 & 70\% vs 85\% & $-.081$ & [-.142, -.019] & $-.512$ & 91.71 & $-2.582$ & .006 & $0.625$ & .267 & Meaningful reduction \\
H5 & 55\% vs 70\% & $-.019$ & [-.070, .032] & $-.148$ & 95.99 & $-0.735$ & .232 & $3.135$ & .001 & No meaningful difference \\
\bottomrule
\end{tabular*}

\raggedright
\smallskip
\footnotesize\textit{Note.} Diff.: mean difference in forward simulation accuracy (lower-correctness $-$ higher-correctness); $d$: Cohen's $d$ (pooled SD); $t_0$, $p_0$: one-tailed test of the mean difference against zero; $t_\Delta$, $p_\Delta$: one-tailed non-inferiority test against the threshold $\Delta = .10$. All tests are one-tailed Welch's \textit{t}-tests, $\alpha = .05$. Degrees of freedom (\textit{df}) are identical for both tests within a comparison, as the non-inferiority test shifts the null value rather than the standard error. Sample sizes: $n_{\text{100\%}} = 51$, $n_{\text{85\%}} = 51$, $n_{\text{70\%}} = 50$, $n_{\text{55\%}} = 48$. Interpretation: Meaningful reduction: significant reduction, effect $\geq \Delta$ not ruled out; Small but detectable: significant reduction, confirmed $< \Delta$; No meaningful difference: no significant reduction, confirmed $< \Delta$; Inconclusive: neither confirmed.
\end{table*}

%% file: figt_forward_sim_accuracy.tex
\begin{figure*}[!tb]
\centering
\includegraphics[width=0.8\textwidth]{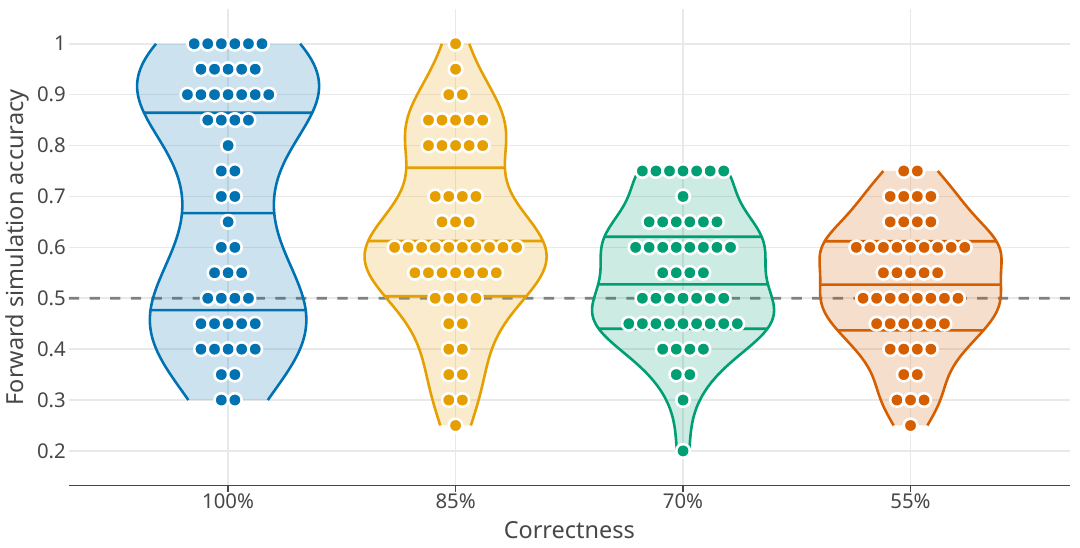}
\caption{Forward simulation accuracy by correctness condition. Each dot represents one participant. Horizontal lines in the violin mark the 25th, 50th, and 75th percentiles; the dashed line marks chance performance (0.5). The 100\% condition is slightly bimodal, with participants clustering at either high accuracy or near chance. The 85\% condition shows a similar high spread but without clear bimodality. In the 70\% and 55\% conditions, variance shrinks and accuracy concentrates near chance (0.5).}
\label{fig:forward-sim-accuracy}
\end{figure*}

%% file: figt_learning_curves.tex
\begin{figure*}[!htbp]
\centering
\includegraphics[width=0.8\textwidth]{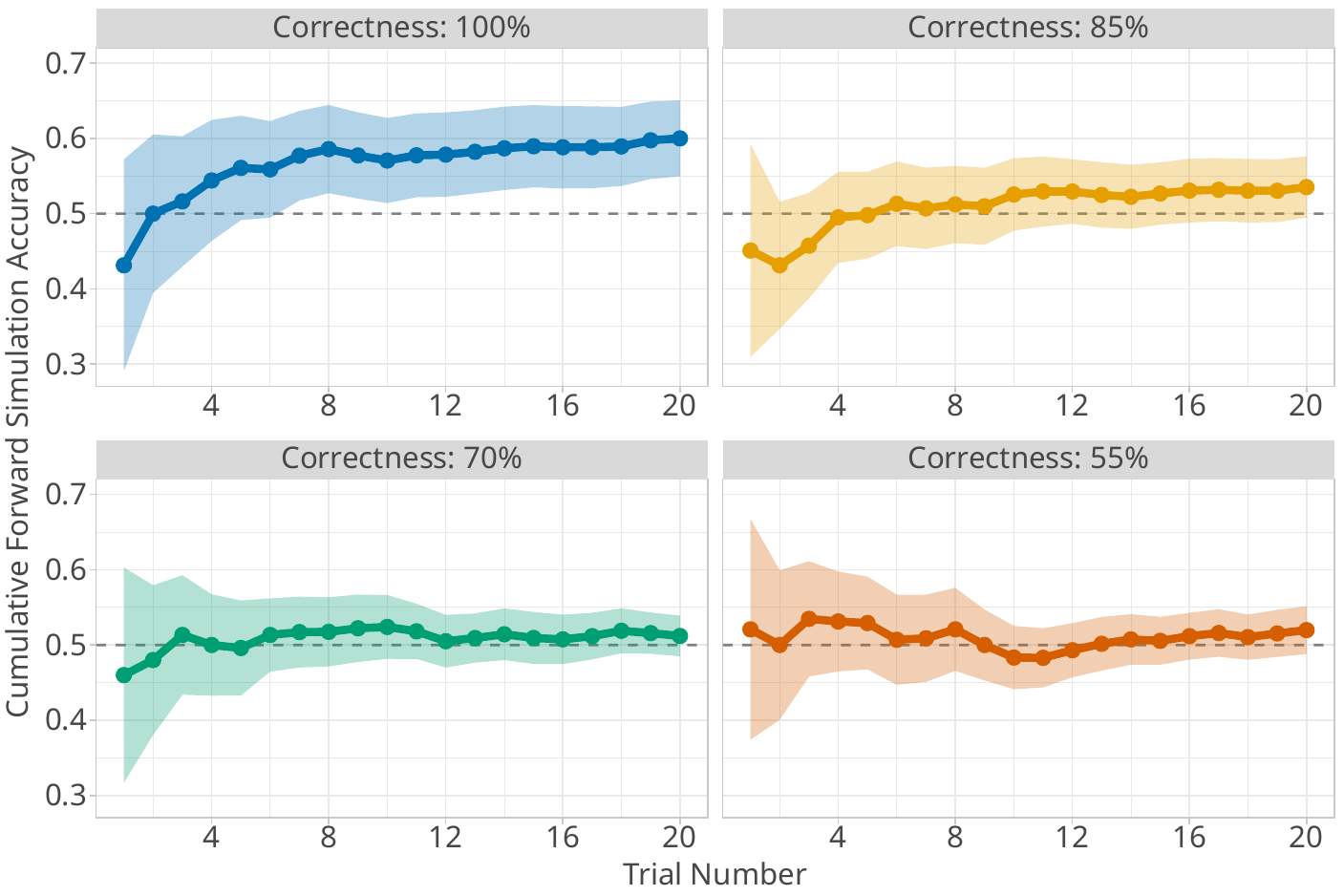}
\caption{Cumulative forward simulation accuracy across training trials by correctness condition, averaged over participants. Shaded ribbons indicate 95\% confidence intervals (CI); the dashed line marks chance performance (0.5). In the 100\% condition, the lower bound of the 95\% CI rises above chance by trial 7 and remains there, suggesting learning. In the remaining conditions, the 95\% CI includes chance throughout, even if the 85\% condition shows slightly higher average accuracy than the 70\% and 55\% conditions toward the end of training.}
\label{fig:learning-curves}
\end{figure*}

%% file: tab_verbalized_rule.tex
\begin{table}[!htbp]
\caption{Verbalized decision rule labels by correctness condition.}
\label{tab:verbalized_rule}
\centering
\small
\setlength{\tabcolsep}{4pt}
\begin{tabular*}{\columnwidth}{@{\extracolsep{\fill}} l r r r r @{}}
\toprule
 & \multicolumn{3}{c}{Verbalized decision rule} & \\
\cmidrule(lr){2-4}
Correctness & Correct & Incorrect & None & Total \\
\midrule
100\% & 14 & 31 & 6 & 51 \\
85\% & 17 & 25 & 9 & 51 \\
70\% & 4 & 37 & 9 & 50 \\
55\% & 2 & 33 & 13 & 48 \\
\midrule
Total & 37 & 126 & 37 & 200 \\
\bottomrule

\end{tabular*}

\raggedright
\smallskip
\footnotesize\textit{Note.} \emph{Correct}: reasoning aligns with the ground-truth rule, referring to localized peaks or troughs. \emph{Incorrect}: participants name other characteristics, such as trend or variance or the location of the highlighted areas. \emph{None}: participants found no pattern or described only the heatmap rather than the underlying data. This measure captures whether participants can verbalize the decision rule, which differs from predicting the AI's output via forward simulation.
\end{table}

%% file: figt_fsim_rule_relationship.tex
\begin{figure*}[!htbp]
\centering
\includegraphics[width=0.8\textwidth]{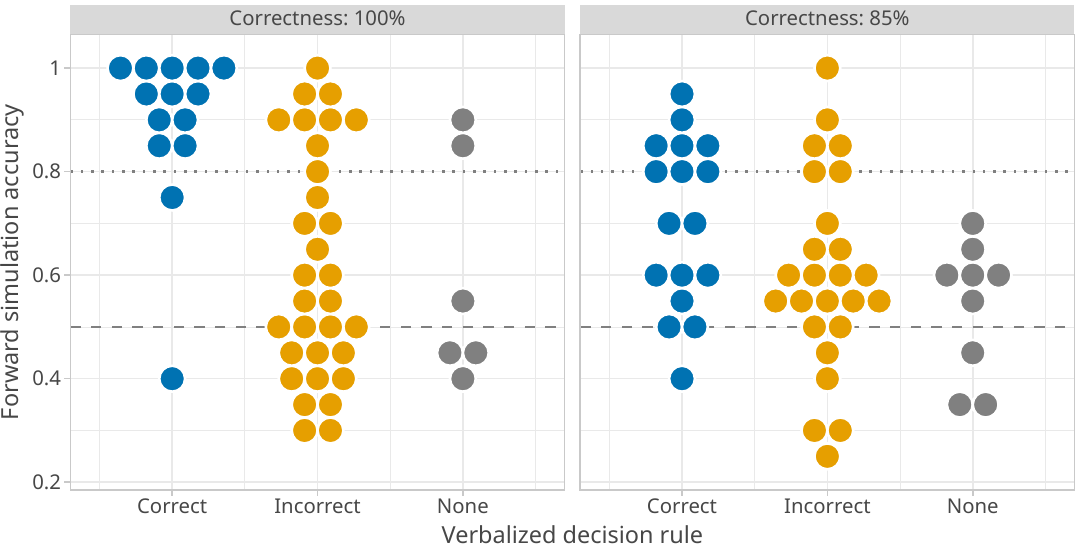}
\caption{Relationship between forward simulation accuracy and the verbalized decision rule (coded as correct, incorrect, or none) for the 100\% and 85\% correct conditions. Each dot is one participant. The dashed line marks chance performance ($.5$) and the dotted line marks a forward simulation accuracy of $.8$, that is, participants predicted 16 or more of the 20 test trials correctly.}
\label{fig:fsim-rule-relationship}
\end{figure*}

%% file: figt_explanation_ratings.tex
\begin{figure*}[!htbp]
\includegraphics[width=\textwidth]{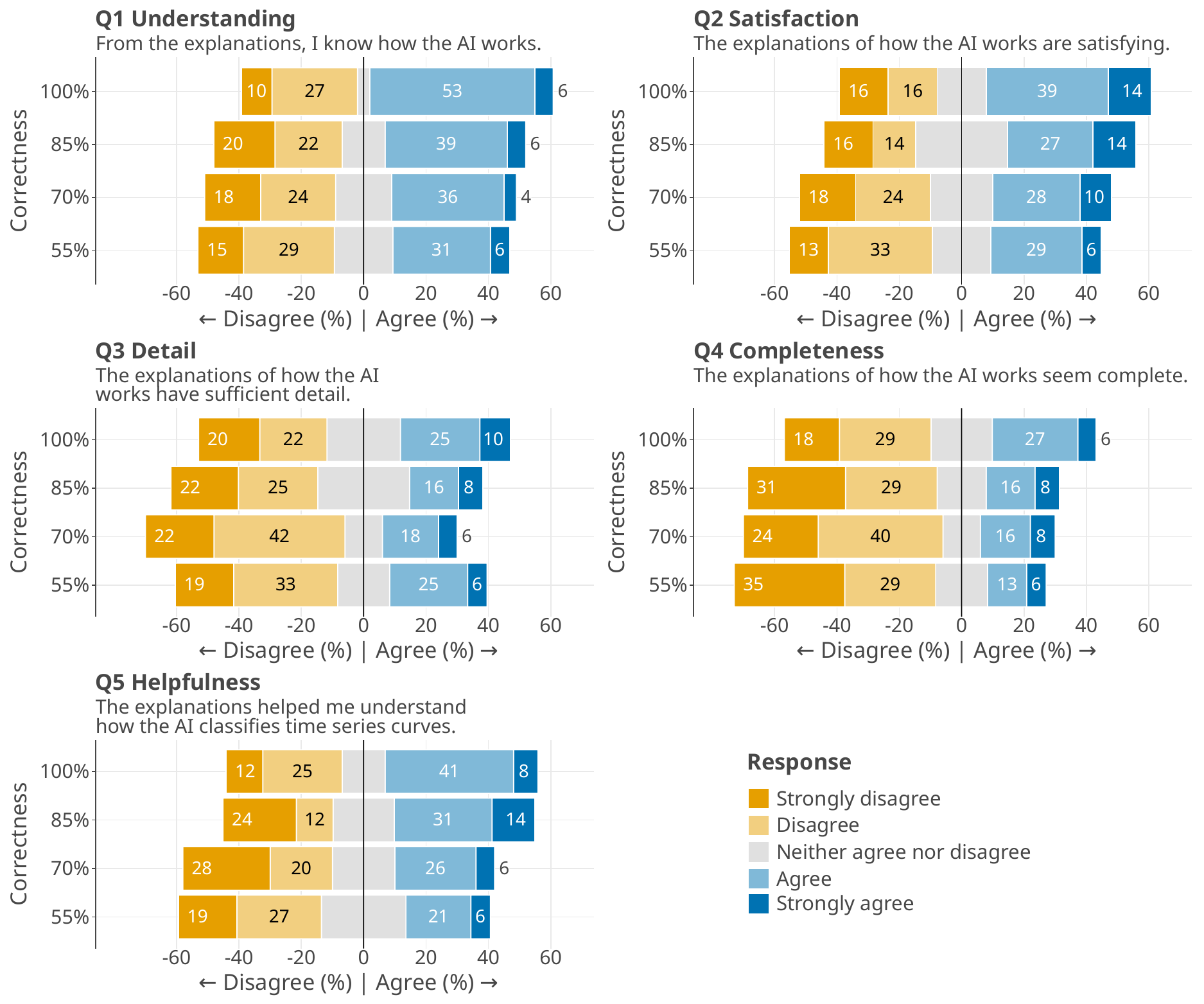}
\caption{Self-reported explanation ratings by correctness condition. Participants rated agreement with five statements adapted from the Explanation Satisfaction Scale on a 5-point Likert scale. Higher correctness conditions show somewhat more positive ratings, but differences across conditions are modest compared to the differences in forward simulation accuracy.}
\label{fig:explanation-ratings}
\end{figure*}

%% file: figt_corr_ess_avg.tex
\begin{figure*}[!tb]
\centering
\includegraphics[width=0.6\textwidth]{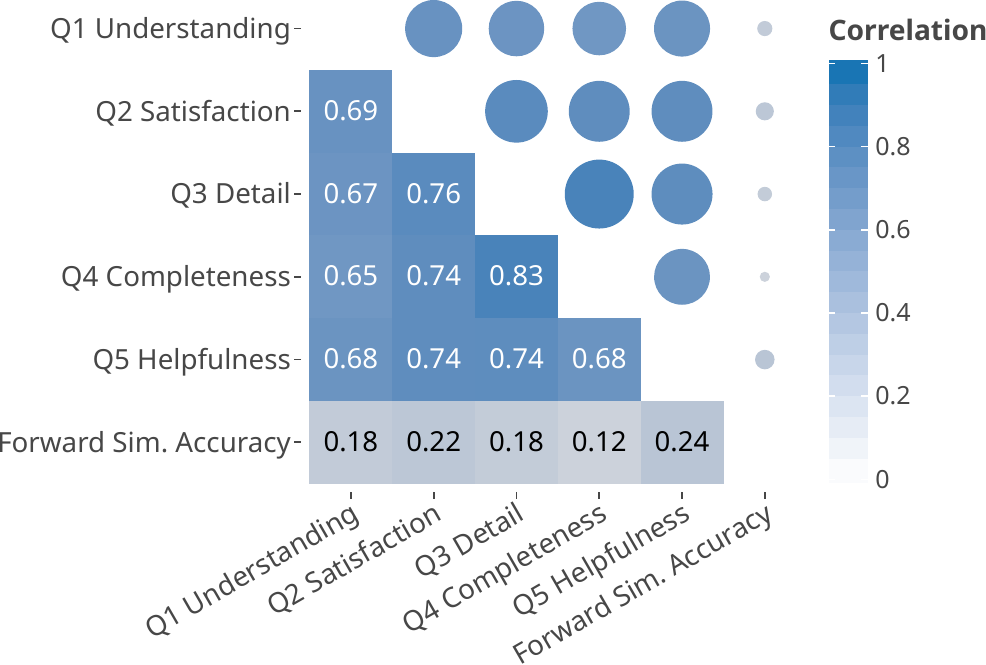}
\caption{Spearman rank correlation between self-reported explanation ratings and forward simulation accuracy, averaged across conditions. The five subjective rating items correlate strongly with each other ($r = .65$ to $.83$) but weakly with forward simulation accuracy ($r \leq .24$).}
\label{fig:correlation-ess-avg}
\end{figure*}

%% file: figt_corr_ess_split.tex
\begin{figure*}[!tb]
\centering
\includegraphics[width=1\textwidth]{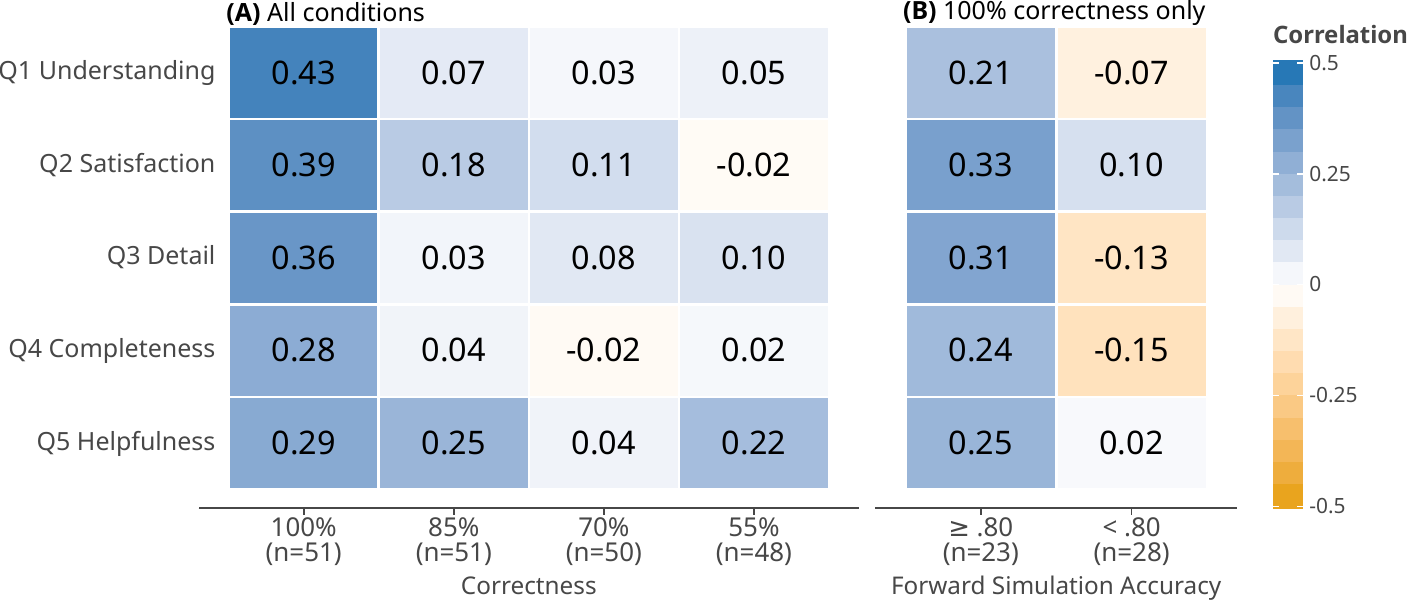}
\caption{Spearman rank correlation between each self-reported explanation rating item and forward simulation accuracy. Each cell shows a single correlation with accuracy. Unlike Figure~\ref{fig:correlation-ess-avg}, inter-item correlations are not shown.
Color encodes correlation on a symmetric $\pm0.5$ scale, set to the observed range ($-0.15$ to $0.43$) rather than the full $\pm1$; cell labels give exact values.
(A)~Split by correctness condition. Correlations were weak to moderate in the 100\% correct condition (up to $r = .43$) and mostly near zero at 85\% and below.
(B)~The 100\% correct condition split at a forward simulation accuracy of $.80$, that is, 16 or more of the 20 test trials correct. At or above the threshold, self-reports correlated weakly to moderately with forward simulation accuracy ($r = .21$ to $.33$); below it, correlations were near zero ($r = -.15$ to $.10$). 
Together, the panels suggest that self-reports corresponded to forward simulation accuracy mainly when explanations were fully correct and participants predicted the AI's decisions accurately.}
\label{fig:correlation-ess-split}
\end{figure*}

%% file: references.bib
@article{spitzer.etal_2025_dont,
  title = {Don't {{Be Fooled}}: {{The Misinformation Effect}} of {{Explanations}} in {{Human}}--{{AI Collaboration}}},
  author = {Spitzer, Philipp and Holstein, Joshua and Morrison, Katelyn and Holstein, Kenneth and Satzger, Gerhard and K{\"u}hl, Niklas},
  year = 2025,
  journal = {International Journal of Human--Computer Interaction},
  volume = {0},
  pages = {1--29},
  doi = {10.1080/10447318.2025.2574511}
}

@inproceedings{piculin.etal_2025_position,
  title = {Position: {{Explainable AI Cannot Advance Without Better User Studies}}},
  booktitle = {Forty-Second {{International Conference}} on {{Machine Learning Position Paper Track}}},
  author = {Pi{\v c}ulin, Matej and Petek, Bernarda and Ograjen{\v s}ek, Irena and Strumbelj, Erik},
  year = 2025,
  url = {https://openreview.net/forum?id=Al5mEX6eHF},
  langid = {english}
}

@article{dwivedi.etal_2023_explainable,
  title = {Explainable {{AI}} ({{XAI}}): {{Core Ideas}}, {{Techniques}}, and {{Solutions}}},
  author = {Dwivedi, Rudresh and Dave, Devam and Naik, Het and Singhal, Smiti and Omer, Rana and Patel, Pankesh and Qian, Bin and Wen, Zhenyu and Shah, Tejal and Morgan, Graham and Ranjan, Rajiv},
  year = 2023,
  journal = {ACM Comput. Surv.},
  volume = {55},
  pages = {194:1--194:33},
  doi = {10.1145/3561048}
}

@article{chanda.etal_2025_dermatologistlike,
  title = {Dermatologist-like Explainable {{AI}} Enhances Melanoma Diagnosis Accuracy: Eye-Tracking Study},
  author = {Chanda, Tirtha and Haggenmueller, Sarah and Bucher, Tabea-Clara and {Holland-Letz}, Tim and Kittler, Harald and Tschandl, Philipp and Heppt, Markus V. and Berking, Carola and Utikal, Jochen S. and Schilling, Bastian and Buerger, Claudia and {Navarrete-Dechent}, Cristian and Goebeler, Matthias and Kather, Jakob Nikolas and Schneider, Carolin V. and Durani, Benjamin and Durani, Hendrike and Jansen, Martin and Wacker, Juliane and Wacker, Joerg and Brinker, Titus J.},
  year = 2025,
  journal = {Nature Communications},
  volume = {16},
  pages = {4739},
  doi = {10.1038/s41467-025-59532-5},
  langid = {english}
}

@article{langer.etal_2021_what,
  title = {What Do We Want from {{Explainable Artificial Intelligence}} ({{XAI}})? -- {{A}} Stakeholder Perspective on {{XAI}} and a Conceptual Model Guiding Interdisciplinary {{XAI}} Research},
  author = {Langer, Markus and Oster, Daniel and Speith, Timo and Hermanns, Holger and K{\"a}stner, Lena and Schmidt, Eva and Sesing, Andreas and Baum, Kevin},
  year = 2021,
  journal = {Artificial Intelligence},
  volume = {296},
  pages = {103473},
  doi = {10.1016/j.artint.2021.103473}
}

@misc{baer_2026_xaitimesynth,
  title = {Xaitimesynth: {{A Python Package}} for {{Evaluating Attribution Methods}} for {{Time Series}} with {{Synthetic Ground Truth}}},
  author = {Baer, Gregor},
  year = 2026,
  doi = {10.48550/arXiv.2603.06781}
}

@inproceedings{bucinca.etal_2025_contrastive,
  title = {Contrastive {{Explanations That Anticipate Human Misconceptions Can Improve Human Decision-Making Skills}}},
  booktitle = {Proceedings of the 2025 {{CHI Conference}} on {{Human Factors}} in {{Computing Systems}}},
  author = {Bu{\c c}inca, Zana and Swaroop, Siddharth and Paluch, Amanda E. and {Doshi-Velez}, Finale and Gajos, Krzysztof Z.},
  year = 2025,
  pages = {1--25},
  doi = {10.1145/3706598.3713229}
}

@article{cheng.etal_2026_uncertainty,
  title = {Uncertainty Explanation of Artificial Intelligence Models by {{SHAP}}},
  author = {Cheng, Ximeng and Wang, Tianqi and Zhu, Di and Ma, Jackie},
  year = 2026,
  journal = {Knowledge-Based Systems},
  volume = {337},
  pages = {115437},
  doi = {10.1016/j.knosys.2026.115437}
}

@article{dejong.etal_2025_cognitive,
  title = {Cognitive {{Forcing}} for {{Better Decision-Making}}: {{Reducing Overreliance}} on {{AI Systems Through Partial Explanations}}},
  author = {De Jong, Sander and Paananen, Ville and Tag, Benjamin and Van Berkel, Niels},
  year = 2025,
  journal = {Proceedings of the ACM on Human-Computer Interaction},
  volume = {9},
  pages = {1--30},
  doi = {10.1145/3710946},
  langid = {english}
}

@inproceedings{bucinca.etal_2020_proxy,
  title = {Proxy Tasks and Subjective Measures Can Be Misleading in Evaluating Explainable {{AI}} Systems},
  booktitle = {Proceedings of the 25th {{International Conference}} on {{Intelligent User Interfaces}}},
  author = {Bu{\c c}inca, Zana and Lin, Phoebe and Gajos, Krzysztof Z. and Glassman, Elena L.},
  year = 2020,
  pages = {454--464},
  doi = {10.1145/3377325.3377498}
}

@inproceedings{cabitza.etal_2024_explanations,
  title = {Explanations {{Considered Harmful}}: {{The Impact}} of~{{Misleading Explanations}} on~{{Accuracy}} in~{{Hybrid Human-AI Decision Making}}},
  booktitle = {Explainable {{Artificial Intelligence}}},
  author = {Cabitza, Federico and Fregosi, Caterina and Campagner, Andrea and Natali, Chiara},
  editor = {Longo, Luca and Lapuschkin, Sebastian and Seifert, Christin},
  year = 2024,
  pages = {255--269},
  doi = {10.1007/978-3-031-63803-9_14},
  langid = {english}
}

@inproceedings{baer.etal_2026_why,
  title = {Why Do {{Class-Dependent Evaluation Effects Occur}} with~{{Time Series Feature Attributions}}? {{A Synthetic Data Investigation}}},
  booktitle = {Machine {{Learning}} and {{Principles}} and {{Practice}} of {{Knowledge Discovery}} in {{Databases}}},
  author = {Baer, Gregor and Grau, Isel and Zhang, Chao and Van Gorp, Pieter},
  editor = {Koprinska, Irena and {Mendes-Moreira}, Jo{\~a}o and Branco, Paula},
  year = 2026,
  pages = {380--395},
  doi = {10.1007/978-3-032-19105-2_27},
  langid = {english}
}

@inproceedings{sadeghi.etal_2024_explaining,
  title = {Explaining the {{Unexplainable}}: {{The Impact}} of {{Misleading Explanations}} on {{Trust}} in {{Unreliable Predictions}} for {{Hardly Assessable Tasks}}},
  booktitle = {Proceedings of the 32nd {{ACM Conference}} on {{User Modeling}}, {{Adaptation}} and {{Personalization}}},
  author = {Sadeghi, Mersedeh and P{\"o}ttgen, Daniel and Ebel, Patrick and Vogelsang, Andreas},
  year = 2024,
  pages = {36--46},
  doi = {10.1145/3627043.3659573}
}

@article{chanda.etal_2024_dermatologistlike,
  title = {Dermatologist-like Explainable {{AI}} Enhances Trust and Confidence in Diagnosing Melanoma},
  author = {Chanda, Tirtha and Hauser, Katja and Hobelsberger, Sarah and Bucher, Tabea-Clara and Garcia, Carina Nogueira and Wies, Christoph and Kittler, Harald and Tschandl, Philipp and {Navarrete-Dechent}, Cristian and Podlipnik, Sebastian and Chousakos, Emmanouil and Crnaric, Iva and Majstorovic, Jovana and Alhajwan, Linda and Foreman, Tanya and Peternel, Sandra and Sarap, Sergei and {\"O}zdemir, {\.I}rem and Barnhill, Raymond L. and {Llamas-Velasco}, Mar and Poch, Gabriela and Korsing, S{\"o}ren and Sondermann, Wiebke and Gellrich, Frank Friedrich and Heppt, Markus V. and Erdmann, Michael and Haferkamp, Sebastian and Drexler, Konstantin and Goebeler, Matthias and Schilling, Bastian and Utikal, Jochen S. and Ghoreschi, Kamran and Fr{\"o}hling, Stefan and {Krieghoff-Henning}, Eva and Brinker, Titus J.},
  year = 2024,
  journal = {Nature Communications},
  volume = {15},
  pages = {524},
  doi = {10.1038/s41467-023-43095-4},
  langid = {english}
}

@inproceedings{adebayo.etal_2018_sanity,
  title = {Sanity {{Checks}} for {{Saliency Maps}}},
  booktitle = {Advances in {{Neural Information Processing Systems}}},
  author = {Adebayo, Julius and Gilmer, Justin and Muelly, Michael and Goodfellow, Ian and Hardt, Moritz and Kim, Been},
  year = 2018,
  volume = {31},
  url = {https://proceedings.neurips.cc/paper_files/paper/2018/hash/294a8ed24b1ad22ec2e7efea049b8737-Abstract.html}
}

@article{samek.etal_2017_evaluating,
  title = {Evaluating the {{Visualization}} of {{What}} a {{Deep Neural Network Has Learned}}},
  author = {Samek, Wojciech and Binder, Alexander and Montavon, Gr{\'e}goire and Lapuschkin, Sebastian and M{\"u}ller, Klaus-Robert},
  year = 2017,
  journal = {IEEE Transactions on Neural Networks and Learning Systems},
  volume = {28},
  pages = {2660--2673},
  doi = {10.1109/TNNLS.2016.2599820}
}

@inproceedings{lakkaraju.bastani_2020_how,
  title = {"{{How}} Do {{I}} Fool You?": {{Manipulating User Trust}} via {{Misleading Black Box Explanations}}},
  booktitle = {Proceedings of the {{AAAI}}/{{ACM Conference}} on {{AI}}, {{Ethics}}, and {{Society}}},
  author = {Lakkaraju, Himabindu and Bastani, Osbert},
  year = 2020,
  pages = {79--85},
  doi = {10.1145/3375627.3375833}
}

@inproceedings{nguyen.ifrim_2025_tshap,
  title = {{{TSHAP}}: {{Fast}} and~{{Exact SHAP}} for~{{Explaining Time Series Classification}} and~{{Regression}}},
  booktitle = {Machine {{Learning}} and {{Knowledge Discovery}} in {{Databases}}. {{Research Track}}},
  author = {Nguyen, Thach Le and Ifrim, Georgiana},
  editor = {Ribeiro, Rita P. and Pfahringer, Bernhard and Japkowicz, Nathalie and Larra{\~n}aga, Pedro and Jorge, Al{\'i}pio M. and Soares, Carlos and Abreu, Pedro H. and Gama, Jo{\~a}o},
  year = 2025,
  pages = {60--77},
  doi = {10.1007/978-3-032-06078-5_4},
  langid = {english}
}

@incollection{kindermans.etal_2019_unreliability,
  title = {The ({{Un}})Reliability of {{Saliency Methods}}},
  booktitle = {Explainable {{AI}}: {{Interpreting}}, {{Explaining}} and {{Visualizing Deep Learning}}},
  author = {Kindermans, Pieter-Jan and Hooker, Sara and Adebayo, Julius and Alber, Maximilian and Sch{\"u}tt, Kristof T. and D{\"a}hne, Sven and Erhan, Dumitru and Kim, Been},
  editor = {Samek, Wojciech and Montavon, Gr{\'e}goire and Vedaldi, Andrea and Hansen, Lars Kai and M{\"u}ller, Klaus-Robert},
  year = 2019,
  pages = {267--280},
  doi = {10.1007/978-3-030-28954-6_14},
  langid = {english}
}

@article{bach.etal_2015_pixelwise,
  title = {On {{Pixel-Wise Explanations}} for {{Non-Linear Classifier Decisions}} by {{Layer-Wise Relevance Propagation}}},
  author = {Bach, Sebastian and Binder, Alexander and Montavon, Gr{\'e}goire and Klauschen, Frederick and M{\"u}ller, Klaus-Robert and Samek, Wojciech},
  year = 2015,
  journal = {PLOS ONE},
  volume = {10},
  pages = {e0130140},
  doi = {10.1371/journal.pone.0130140},
  langid = {english}
}

@article{arras.etal_2022_clevrxai,
  title = {{{CLEVR-XAI}}: {{A}} Benchmark Dataset for the Ground Truth Evaluation of Neural Network Explanations},
  author = {Arras, Leila and Osman, Ahmed and Samek, Wojciech},
  year = 2022,
  journal = {Information Fusion},
  volume = {81},
  pages = {14--40},
  doi = {10.1016/j.inffus.2021.11.008}
}

@article{morrison.etal_2024_impact,
  title = {The {{Impact}} of {{Imperfect XAI}} on {{Human-AI Decision-Making}}},
  author = {Morrison, Katelyn and Spitzer, Philipp and Turri, Violet and Feng, Michelle and K{\"u}hl, Niklas and Perer, Adam},
  year = 2024,
  journal = {Proc. ACM Hum.-Comput. Interact.},
  volume = {8},
  pages = {183:1--183:39},
  doi = {10.1145/3641022}
}

@article{chen.etal_2023_machine,
  title = {Machine {{Explanations}} and {{Human Understanding}}},
  author = {Chen, Chacha and Feng, Shi and Sharma, Amit and Tan, Chenhao},
  year = 2023,
  journal = {Transactions on Machine Learning Research},
  url = {https://openreview.net/forum?id=y4CGF1A8VG},
  langid = {english}
}

@inproceedings{fong.vedaldi_2017_interpretable,
  title = {Interpretable {{Explanations}} of {{Black Boxes}} by {{Meaningful Perturbation}}},
  booktitle = {Proceedings of the {{IEEE International Conference}} on {{Computer Vision}}},
  author = {Fong, Ruth C. and Vedaldi, Andrea},
  year = 2017,
  pages = {3429--3437},
  url = {https://openaccess.thecvf.com/content_iccv_2017/html/Fong_Interpretable_Explanations_of_ICCV_2017_paper.html}
}

@inproceedings{alqaraawi.etal_2020_evaluating,
  title = {Evaluating Saliency Map Explanations for Convolutional Neural Networks: A User Study},
  booktitle = {Proceedings of the 25th {{International Conference}} on {{Intelligent User Interfaces}}},
  author = {Alqaraawi, Ahmed and Schuessler, Martin and Wei{\ss}, Philipp and Costanza, Enrico and Berthouze, Nadia},
  year = 2020,
  pages = {275--285},
  doi = {10.1145/3377325.3377519}
}

@article{longo.etal_2024_explainable,
  title = {Explainable {{Artificial Intelligence}} ({{XAI}}) 2.0: {{A}} Manifesto of Open Challenges and Interdisciplinary Research Directions},
  author = {Longo, Luca and Brcic, Mario and Cabitza, Federico and Choi, Jaesik and Confalonieri, Roberto and Ser, Javier Del and Guidotti, Riccardo and Hayashi, Yoichi and Herrera, Francisco and Holzinger, Andreas and Jiang, Richard and Khosravi, Hassan and Lecue, Freddy and Malgieri, Gianclaudio and P{\'a}ez, Andr{\'e}s and Samek, Wojciech and Schneider, Johannes and Speith, Timo and Stumpf, Simone},
  year = 2024,
  journal = {Information Fusion},
  volume = {106},
  pages = {102301},
  doi = {10.1016/j.inffus.2024.102301}
}

@article{vasconcelos.etal_2023_explanations,
  title = {Explanations {{Can Reduce Overreliance}} on {{AI Systems During Decision-Making}}},
  author = {Vasconcelos, Helena and J{\"o}rke, Matthew and {Grunde-McLaughlin}, Madeleine and Gerstenberg, Tobias and Bernstein, Michael S. and Krishna, Ranjay},
  year = 2023,
  journal = {Proc. ACM Hum.-Comput. Interact.},
  volume = {7},
  pages = {129:1--129:38},
  doi = {10.1145/3579605}
}

@inproceedings{petsiuk.etal_2018_rise,
  title = {{{RISE}}: {{Randomized Input Sampling}} for {{Explanation}} of {{Black}}-Box {{Models}}},
  booktitle = {Proceedings of the {{BMVC}}},
  author = {Petsiuk, Vitali and Das, Abir and Saenko, Kate},
  year = 2018,
}

@inproceedings{kaur.etal_2020_interpreting,
  title = {Interpreting {{Interpretability}}: {{Understanding Data Scientists}}' {{Use}} of {{Interpretability Tools}} for {{Machine Learning}}},
  booktitle = {Proceedings of the 2020 {{CHI Conference}} on {{Human Factors}} in {{Computing Systems}}},
  author = {Kaur, Harmanpreet and Nori, Harsha and Jenkins, Samuel and Caruana, Rich and Wallach, Hanna and Wortman Vaughan, Jennifer},
  year = 2020,
  pages = {1--14},
  doi = {10.1145/3313831.3376219}
}

@article{lipton_2018_mythos,
  title = {The {{Mythos}} of {{Model Interpretability}}: {{In}} Machine Learning, the Concept of Interpretability Is Both Important and Slippery.},
  author = {Lipton, Zachary C.},
  year = 2018,
  journal = {Queue},
  volume = {16},
  pages = {31--57},
  doi = {10.1145/3236386.3241340},
  langid = {english}
}

@inproceedings{kim.etal_2022_hive,
  title = {{{HIVE}}: {{Evaluating}} the {{Human Interpretability}} of {{Visual Explanations}}},
  booktitle = {Computer {{Vision}} -- {{ECCV}} 2022},
  author = {Kim, Sunnie S. Y. and Meister, Nicole and Ramaswamy, Vikram V. and Fong, Ruth and Russakovsky, Olga},
  editor = {Avidan, Shai and Brostow, Gabriel and Ciss{\'e}, Moustapha and Farinella, Giovanni Maria and Hassner, Tal},
  year = 2022,
  pages = {280--298},
  doi = {10.1007/978-3-031-19775-8_17},
  langid = {english}
}

@inproceedings{speith_2026_review,
  title = {A {{Review}} of {{Taxonomies}} of {{Explainable AI}} ({{XAI}}) {{Evaluation}}},
  booktitle = {Proceedings of the 2026 {{ACM Conference}} on {{Fairness}}, {{Accountability}}, and {{Transparency}}},
  author = {Speith, Timo},
  year = 2026,
  pages = {541--569},
  doi = {10.1145/3805689.3806537}
}

@article{speith.etal_2024_conceptualizing,
  title = {Conceptualizing Understanding in Explainable Artificial Intelligence ({{XAI}}): An Abilities-Based Approach},
  author = {Speith, Timo and Crook, Barnaby and Mann, Sara and Schom{\"a}cker, Astrid and Langer, Markus},
  year = 2024,
  journal = {Ethics and Information Technology},
  volume = {26},
  pages = {40},
  doi = {10.1007/s10676-024-09769-3},
  langid = {english}
}

@inproceedings{hase.bansal_2020_evaluating,
  title = {Evaluating {{Explainable AI}}: {{Which Algorithmic Explanations Help Users Predict Model Behavior}}?},
  booktitle = {Proceedings of the 58th {{Annual Meeting}} of the {{Association}} for {{Computational Linguistics}}},
  author = {Hase, Peter and Bansal, Mohit},
  editor = {Jurafsky, Dan and Chai, Joyce and Schluter, Natalie and Tetreault, Joel},
  year = 2020,
  pages = {5540--5552},
  doi = {10.18653/v1/2020.acl-main.491}
}

@article{khan.etal_2024_faithful,
  title = {Faithful {{Counterfactual Visual Explanations}} ({{FCVE}})},
  author = {Khan, Bismillah and Tariq, Syed Ali and Zia, Tehseen and Ahsan, Muhammad and Windridge, David},
  year = 2024,
  journal = {Knowledge-Based Systems},
  volume = {294},
  pages = {111668},
  doi = {10.1016/j.knosys.2024.111668}
}

@article{simic.etal_2025_comprehensive,
  title = {A Comprehensive Analysis of Perturbation Methods in Explainable {{AI}} Feature Attribution Validation for Neural Time Series Classifiers},
  author = {{\v S}imi{\'c}, Ilija and Veas, Eduardo and Sabol, Vedran},
  year = 2025,
  journal = {Scientific Reports},
  volume = {15},
  pages = {26607},
  doi = {10.1038/s41598-025-09538-2},
  langid = {english}
}

@article{kares.etal_2026_are,
  title = {Are Explainable {{AI}} ({{XAI}}) Evaluation Strategies Aligned? {{Comparing}} Subjective, Objective, and Mathematical Evaluation Measures Using Saliency Maps},
  author = {Kares, Felix and Speith, Timo and Zhang, Hanwei and Langer, Markus},
  year = 2026,
  journal = {Computers in Human Behavior},
  pages = {109163},
  doi = {10.1016/j.chb.2026.109163}
}

@inproceedings{simonyan.etal_2014_deep,
  title = {Deep inside Convolutional Networks: Visualising Image Classification Models and Saliency Maps},
  booktitle = {Proceedings of the {{International Conference}} on {{Learning Representations}} ({{ICLR}})},
  author = {Simonyan, K and Vedaldi, A and Zisserman, A},
  year = 2014,
  organizer = {2nd International Conference on Learning Representations (ICLR 2014)}
}

@article{bluecher.etal_2024_decoupling,
  title = {Decoupling {{Pixel Flipping}} and {{Occlusion Strategy}} for {{Consistent XAI Benchmarks}}},
  author = {Bluecher, Stefan and Vielhaben, Johanna and Strodthoff, Nils},
  year = 2024,
  journal = {Transactions on Machine Learning Research},
  url = {https://openreview.net/forum?id=bIiLXdtUVM},
  langid = {english}
}

@inproceedings{arias-duart.etal_2022_focus,
  title = {Focus! {{Rating XAI Methods}} and {{Finding Biases}}},
  booktitle = {2022 {{IEEE International Conference}} on {{Fuzzy Systems}} ({{FUZZ-IEEE}})},
  author = {{Arias-Duart}, Anna and Par{\'e}s, Ferran and {Garcia-Gasulla}, Dario and {Gim{\'e}nez-{\'A}balos}, Victor},
  year = 2022,
  pages = {1--8},
  doi = {10.1109/FUZZ-IEEE55066.2022.9882821}
}

@inproceedings{suh.etal_2025_fewer,
  title = {Fewer {{Than}} 1\% of {{Explainable AI Papers Validate Explainability}} with {{Humans}}},
  booktitle = {Proceedings of the {{Extended Abstracts}} of the {{CHI Conference}} on {{Human Factors}} in {{Computing Systems}}},
  author = {Suh, Ashley and Hurley, Isabelle and Smith, Nora and Siu, Ho Chit},
  year = 2025,
  pages = {1--7},
  doi = {10.1145/3706599.3719964}
}

@inproceedings{zeiler.fergus_2014_visualizing,
  title = {Visualizing and {{Understanding Convolutional Networks}}},
  booktitle = {Computer {{Vision}} -- {{ECCV}} 2014},
  author = {Zeiler, Matthew D. and Fergus, Rob},
  editor = {Fleet, David and Pajdla, Tomas and Schiele, Bernt and Tuytelaars, Tinne},
  year = 2014,
  pages = {818--833},
  doi = {10.1007/978-3-319-10590-1_53},
  langid = {english}
}

@article{kenny.etal_2021_explaining,
  title = {Explaining Black-Box Classifiers Using {\emph{Post-Hoc}} Explanations-by-Example: {{The}} Effect of Explanations and Error-Rates in {{XAI}} User Studies},
  author = {Kenny, Eoin M. and Ford, Courtney and Quinn, Molly and Keane, Mark T.},
  year = 2021,
  journal = {Artificial Intelligence},
  volume = {294},
  pages = {103459},
  doi = {10.1016/j.artint.2021.103459}
}

@article{rong.etal_2024_humancentered,
  title = {Towards {{Human-Centered Explainable AI}}: {{A Survey}} of {{User Studies}} for {{Model Explanations}}},
  author = {Rong, Yao and Leemann, Tobias and Nguyen, Thai-Trang and Fiedler, Lisa and Qian, Peizhu and Unhelkar, Vaibhav and Seidel, Tina and Kasneci, Gjergji and Kasneci, Enkelejda},
  year = 2024,
  journal = {IEEE Transactions on Pattern Analysis and Machine Intelligence},
  volume = {46},
  pages = {2104--2122},
  doi = {10.1109/TPAMI.2023.3331846}
}

@inproceedings{abbaspouronari.etal_2025_dynamics,
  title = {The {{Dynamics}} of~{{Trust}} in~{{XAI}}: {{Assessing Perceived}} and~{{Demonstrated Trust Across Interaction Modes}} and~{{Risk Treatments}}},
  booktitle = {Explainable {{Artificial Intelligence}}},
  author = {Abbaspour Onari, Mohsen and Baer, Gregor and Zhang, Chao and Grau, Isel and Nobile, Marco S. and Zhang, Yingqian},
  editor = {Guidotti, Riccardo and Schmid, Ute and Longo, Luca},
  year = 2025,
  pages = {316--334},
  doi = {10.1007/978-3-032-08317-3_15},
  langid = {english}
}

@inproceedings{baer.etal_2025_classdependent,
  title = {Class-{{Dependent Perturbation Effects}} in~{{Evaluating Time Series Attributions}}},
  booktitle = {Explainable {{Artificial Intelligence}}},
  author = {Baer, Gregor and Grau, Isel and Zhang, Chao and Van Gorp, Pieter},
  editor = {Guidotti, Riccardo and Schmid, Ute and Longo, Luca},
  year = 2025,
  pages = {292--314},
  doi = {10.1007/978-3-032-08330-2_14},
  langid = {english}
}

@article{zhang.etal_2018_topdown,
  title = {Top-{{Down Neural Attention}} by {{Excitation Backprop}}},
  author = {Zhang, Jianming and Bargal, Sarah Adel and Lin, Zhe and Brandt, Jonathan and Shen, Xiaohui and Sclaroff, Stan},
  year = 2018,
  journal = {International Journal of Computer Vision},
  volume = {126},
  pages = {1084--1102},
  doi = {10.1007/s11263-017-1059-x},
  langid = {english}
}

@article{ismail.etal_2020_benchmarking,
  title = {Benchmarking Deep Learning Interpretability in Time Series Predictions},
  author = {Ismail, Aya Abdelsalam and Gunady, Mohamed and Corrada Bravo, Hector and Feizi, Soheil},
  year = 2020,
  journal = {Advances in neural information processing systems},
  volume = {33},
  pages = {6441--6452},
  url = {https://proceedings.neurips.cc/paper_files/paper/2020/hash/47a3893cc405396a5c30d91320572d6d-Abstract.html}
}

@incollection{doshi-velez.kim_2018_considerations,
  title = {Considerations for {{Evaluation}} and {{Generalization}} in {{Interpretable Machine Learning}}},
  booktitle = {Explainable and {{Interpretable Models}} in {{Computer Vision}} and {{Machine Learning}}},
  author = {{Doshi-Velez}, Finale and Kim, Been},
  editor = {Escalante, Hugo Jair and Escalera, Sergio and Guyon, Isabelle and Bar{\'o}, Xavier and G{\"u}{\c c}l{\"u}t{\"u}rk, Ya{\u g}mur and G{\"u}{\c c}l{\"u}, Umut and {van Gerven}, Marcel},
  year = 2018,
  pages = {3--17},
  doi = {10.1007/978-3-319-98131-4_1},
  langid = {english}
}

@article{turbe.etal_2023_evaluation,
  title = {Evaluation of Post-Hoc Interpretability Methods in Time-Series Classification},
  author = {Turb{\'e}, Hugues and Bjelogrlic, Mina and Lovis, Christian and Mengaldo, Gianmarco},
  year = 2023,
  journal = {Nature Machine Intelligence},
  volume = {5},
  pages = {250--260},
  doi = {10.1038/s42256-023-00620-w},
  langid = {english}
}

@inproceedings{schlegel.etal_2019_rigorous,
  title = {Towards a Rigorous Evaluation of {{XAI}} Methods on Time Series},
  booktitle = {2019 {{IEEE}}/{{CVF}} International Conference on Computer Vision Workshop ({{ICCVW}})},
  author = {Schlegel, Udo and Arnout, Hiba and {El-Assady}, Mennatallah and Oelke, Daniela and Keim, Daniel A.},
  year = 2019,
  pages = {4197--4201},
  doi = {10.1109/ICCVW.2019.00516}
}

@inproceedings{lundberg.lee_2017_unified,
  title = {A {{Unified Approach}} to {{Interpreting Model Predictions}}},
  booktitle = {Advances in {{Neural Information Processing Systems}}},
  author = {Lundberg, Scott M and Lee, Su-In},
  editor = {Guyon, I. and Luxburg, U. Von and Bengio, S. and Wallach, H. and Fergus, R. and Vishwanathan, S. and Garnett, R.},
  year = 2017,
  volume = {30},
  url = {https://proceedings.neurips.cc/paper_files/paper/2017/file/8a20a8621978632d76c43dfd28b67767-Paper.pdf}
}

@article{henninger.etal_2022_labjs,
  title = {Lab.Js: {{A}} Free, Open, Online Study Builder},
  author = {Henninger, Felix and Shevchenko, Yury and Mertens, Ulf K. and Kieslich, Pascal J. and Hilbig, Benjamin E.},
  year = 2022,
  journal = {Behavior Research Methods},
  volume = {54},
  pages = {556--573},
  doi = {10.3758/s13428-019-01283-5},
  langid = {english}
}

@article{gruensteidl.kirrane_2026_evaluation,
  title = {An {{Evaluation}} of {{XAI Methods}} Applied to {{Information Disorder Detection Models}}},
  author = {Gruensteidl, Mark Nicolas and Kirrane, Sabrina},
  year = 2026,
  journal = {Knowledge-Based Systems},
  volume = {336},
  pages = {115305},
  doi = {10.1016/j.knosys.2026.115305}
}

@inproceedings{wang.etal_2024_impact,
  title = {Impact of~{{Fidelity}} and~{{Robustness}} of~{{Machine Learning Explanations}} on~{{User Trust}}},
  booktitle = {{{AI}} 2023: {{Advances}} in {{Artificial Intelligence}}},
  author = {Wang, Bo and Zhou, Jianlong and Li, Yiqiao and Chen, Fang},
  editor = {Liu, Tongliang and Webb, Geoff and Yue, Lin and Wang, Dadong},
  year = 2024,
  pages = {209--220},
  doi = {10.1007/978-981-99-8391-9_17},
  langid = {english}
}

@article{hedstrom.etal_2023_metaevaluation,
  title = {The {{Meta-Evaluation Problem}} in {{Explainable AI}}: {{Identifying Reliable Estimators}} with {{MetaQuantus}}},
  author = {Hedstr{\"o}m, Anna and Bommer, Philine Lou and Wickstr{\o}m, Kristoffer Knutsen and Samek, Wojciech and Lapuschkin, Sebastian and H{\"o}hne, Marina MC},
  year = 2023,
  journal = {Transactions on Machine Learning Research},
  url = {https://openreview.net/forum?id=j3FK00HyfU},
  langid = {english}
}

@incollection{schlegel.keim_2023_deep,
  title = {A {{Deep Dive}} into {{Perturbations}} as {{Evaluation Technique}} for {{Time Series XAI}}},
  booktitle = {Explainable {{Artificial Intelligence}}},
  author = {Schlegel, Udo and Keim, Daniel A.},
  editor = {Longo, Luca},
  year = 2023,
  volume = {1903},
  pages = {165--180},
  doi = {10.1007/978-3-031-44070-0_9},
  langid = {english}
}

@inproceedings{schulz.etal_2020_restricting,
  title = {Restricting the Flow: {{Information}} Bottlenecks for Attribution},
  booktitle = {International Conference on Learning Representations},
  author = {Schulz, Karl and Sixt, Leon and Tombari, Federico and Landgraf, Tim},
  year = 2020,
  url = {https://openreview.net/forum?id=S1xWh1rYwB}
}

@inproceedings{papenmeier.etal_2019_how,
  title = {How Model Accuracy and Explanation Fidelity Influence User Trust in {{AI}}},
  booktitle = {{{IJCAI Workshop}} on {{Explainable Artificial Intelligence}} ({{XAI}}) 2019},
  author = {Papenmeier, Andrea and Englebienne, Gwenn and Seifert, Christin},
  year = 2019,
  url = {https://research.utwente.nl/files/480379145/Papenmeier2019_xai_trust_influence_preprint.pdf}
}

@misc{smilkov.etal_2017_smoothgrad,
  title = {{{SmoothGrad}}: Removing Noise by Adding Noise},
  author = {Smilkov, Daniel and Thorat, Nikhil and Kim, Been and Vi{\'e}gas, Fernanda and Wattenberg, Martin},
  year = 2017,
  doi = {10.48550/arXiv.1706.03825}
}

@article{nauta.etal_2023_anecdotal,
  title = {From {{Anecdotal Evidence}} to {{Quantitative Evaluation Methods}}: {{A Systematic Review}} on {{Evaluating Explainable AI}}},
  author = {Nauta, Meike and Trienes, Jan and Pathak, Shreyasi and Nguyen, Elisa and Peters, Michelle and Schmitt, Yasmin and Schl{\"o}tterer, J{\"o}rg and Van Keulen, Maurice and Seifert, Christin},
  year = 2023,
  journal = {ACM Computing Surveys},
  volume = {55},
  pages = {1--42},
  doi = {10.1145/3583558},
  langid = {english}
}

@article{guidotti.etal_2018_survey,
  title = {A {{Survey}} of {{Methods}} for {{Explaining Black Box Models}}},
  author = {Guidotti, Riccardo and Monreale, Anna and Ruggieri, Salvatore and Turini, Franco and Giannotti, Fosca and Pedreschi, Dino},
  year = 2018,
  journal = {ACM Comput. Surv.},
  volume = {51},
  pages = {93:1--93:42},
  doi = {10.1145/3236009}
}

@article{krishna.etal_2024_disagreement,
  title = {The {{Disagreement Problem}} in {{Explainable Machine Learning}}: {{A Practitioner}}'s {{Perspective}}},
  author = {Krishna, Satyapriya and Han, Tessa and Gu, Alex and Wu, Steven and Jabbari, Shahin and Lakkaraju, Himabindu},
  year = 2024,
  journal = {Transactions on Machine Learning Research},
  url = {https://openreview.net/forum?id=jESY2WTZCe},
  langid = {english}
}

@inproceedings{bansal.etal_2021_does,
  title = {Does the {{Whole Exceed}} Its {{Parts}}? {{The Effect}} of {{AI Explanations}} on {{Complementary Team Performance}}},
  booktitle = {Proceedings of the 2021 {{CHI Conference}} on {{Human Factors}} in {{Computing Systems}}},
  author = {Bansal, Gagan and Wu, Tongshuang and Zhou, Joyce and Fok, Raymond and Nushi, Besmira and Kamar, Ece and Ribeiro, Marco Tulio and Weld, Daniel},
  year = 2021,
  pages = {1--16},
  doi = {10.1145/3411764.3445717}
}

@inproceedings{shrikumar.etal_2017_learning,
  title = {Learning {{Important Features Through Propagating Activation Differences}}},
  booktitle = {Proceedings of the 34th {{International Conference}} on {{Machine Learning}}},
  author = {Shrikumar, Avanti and Greenside, Peyton and Kundaje, Anshul},
  year = 2017,
  pages = {3145--3153},
  url = {https://proceedings.mlr.press/v70/shrikumar17a.html},
  langid = {english}
}

@inproceedings{sundararajan.etal_2017_axiomatic,
  title = {Axiomatic {{Attribution}} for {{Deep Networks}}},
  booktitle = {Proceedings of the 34th {{International Conference}} on {{Machine Learning}}},
  author = {Sundararajan, Mukund and Taly, Ankur and Yan, Qiqi},
  year = 2017,
  pages = {3319--3328},
  url = {https://proceedings.mlr.press/v70/sundararajan17a.html},
  langid = {english}
}

@inproceedings{chromik.etal_2021_think,
  title = {I {{Think I Get Your Point}}, {{AI}}! {{The Illusion}} of {{Explanatory Depth}} in {{Explainable AI}}},
  booktitle = {Proceedings of the 26th {{International Conference}} on {{Intelligent User Interfaces}}},
  author = {Chromik, Michael and Eiband, Malin and Buchner, Felicitas and Kr{\"u}ger, Adrian and Butz, Andreas},
  year = 2021,
  pages = {307--317},
  doi = {10.1145/3397481.3450644}
}

@inproceedings{rawal.etal_2025_evaluating,
  title = {Evaluating {{Model Explanations}} without {{Ground Truth}}},
  booktitle = {Proceedings of the 2025 {{ACM Conference}} on {{Fairness}}, {{Accountability}}, and {{Transparency}}},
  author = {Rawal, Kaivalya and Fu, Zihao and Delaney, Eoin and Russell, Chris},
  year = 2025,
  pages = {3400--3411},
  doi = {10.1145/3715275.3732219}
}

@article{hoffman.etal_2023_measures,
  title = {Measures for Explainable {{AI}}: {{Explanation}} Goodness, User Satisfaction, Mental Models, Curiosity, Trust, and Human-{{AI}} Performance},
  author = {Hoffman, Robert R. and Mueller, Shane T. and Klein, Gary and Litman, Jordan},
  year = 2023,
  journal = {Frontiers in Computer Science},
  volume = {5},
  pages = {1096257},
  url = {https://www.frontiersin.org/articles/10.3389/fcomp.2023.1096257/full}
}

@inproceedings{hooker.etal_2019_benchmark,
  title = {A {{Benchmark}} for {{Interpretability Methods}} in {{Deep Neural Networks}}},
  booktitle = {Advances in {{Neural Information Processing Systems}}},
  author = {Hooker, Sara and Erhan, Dumitru and Kindermans, Pieter-Jan and Kim, Been},
  year = 2019,
  volume = {32},
  url = {https://proceedings.neurips.cc/paper_files/paper/2019/hash/fe4b8556000d0f0cae99daa5c5c5a410-Abstract.html}
}

@article{nguyen.etal_2021_effectiveness,
  title = {The Effectiveness of Feature Attribution Methods and Its Correlation with Automatic Evaluation Scores},
  author = {Nguyen, Giang and Kim, Daeyoung and Nguyen, Anh},
  year = 2021,
  journal = {Advances in Neural Information Processing Systems},
  volume = {34},
  pages = {26422--26436},
  url = {https://proceedings.neurips.cc/paper/2021/hash/de043a5e421240eb846da8effe472ff1-Abstract.html}
}

@article{colin.etal_2022_what,
  title = {What {{I}} Cannot Predict, {{I}} Do Not Understand: {{A}} Human-Centered Evaluation Framework for Explainability Methods},
  author = {Colin, Julien and Fel, Thomas and Cad{\`e}ne, R{\'e}mi and Serre, Thomas},
  year = 2022,
  journal = {Advances in neural information processing systems},
  volume = {35},
  pages = {2832--2845},
  url = {https://proceedings.neurips.cc/paper_files/paper/2022/hash/13113e938f2957891c0c5e8df811dd01-Abstract-Conference.html}
}

@misc{ribeiro.etal_2016_why,
  title = {"{{Why Should I Trust You}}?": {{Explaining}} the {{Predictions}} of {{Any Classifier}}},
  author = {Ribeiro, Marco Tulio and Singh, Sameer and Guestrin, Carlos},
  year = 2016,
  langid = {english},
  doi = {10.48550/arXiv.1602.04938}
}

@incollection{grau.napoles_2024_sparsenessoptimized,
  title = {Sparseness-{{Optimized Feature Importance}}},
  booktitle = {Explainable {{Artificial Intelligence}}},
  author = {Grau, Isel and N{\'a}poles, Gonzalo},
  editor = {Longo, Luca and Lapuschkin, Sebastian and Seifert, Christin},
  year = 2024,
  volume = {2154},
  pages = {393--415},
  doi = {10.1007/978-3-031-63797-1_20},
  langid = {english}
}
